\documentclass[aps,prb,preprint,superscriptaddress,showpacs]{revtex4-1}

\usepackage{textcomp}
\usepackage{amsmath}
\usepackage{amssymb}
\usepackage{bm}
\usepackage{graphicx}
\usepackage{color}
\usepackage{dcolumn} 
\usepackage[caption=false]{subfig} 

\newcommand{\rvec}{\textbf{r}}

\newcommand{\bigO}{O} 
\newcommand{\mean}[1]{\overline{#1}} 

\newcommand{\jmmapprox}{\sim \! \!}

\newcolumntype{d}[1]{D{.}{.}{#1}}
\newcolumntype{e}[1]{D{.}{}{#1}}

\begin{document}

\title{Density functionals from deep learning}

\author{Jeffrey M.\ McMahon}
\email[]{jeffrey.mcmahon@wsu.edu}
\affiliation{Department of Physics and Astronomy, Washington State University, Pullman, Washington 99164, USA}

\date{\today}


\begin{abstract}

Density-functional theory is a formally exact description of a many-body quantum system in terms of its density; in practice, however, approximations to the universal density functional are required. In this work, a model based on deep learning is developed to approximate this functional. Deep learning allows computational models that are capable of naturally discovering intricate structure in large and/or high-dimensional data sets, with multiple levels of abstraction. As no assumptions are made as to the form of this structure, this approach is much more powerful and flexible than traditional approaches. As an example application, the model is shown to perform well on approximating the kinetic-energy density functional for noninteracting electrons. The model is analyzed in detail, and its advantages over conventional machine learning are discussed. 

\end{abstract}





\maketitle



\section{Introduction}
\label{sec:intro}

Given any system of interacting electrons in an external potential $v(\rvec)$, the theorems of Hohenberg and Kohn \cite{PhysRev.136.B864} prove that there exists a universal (though unknown) functional of the density $n$, $F[n]$, independent of $v(\rvec)$, such that the expression:
\begin{equation}
    \label{eq:HK}
    E[n] = \int d\rvec ~ v(\rvec) n(\rvec) + F[n] 
\end{equation}
is minimized and equal to the ground-state energy when $n$ is equal to the ground-state density. After separating out the classical Coulomb (Hartree) energy $G[n] = F[n] - E_\text{H}[n]$, Kohn and Sham \cite{PhysRev.140.A1133} developed a general method (as follows) to find the solution to Eq.\ (\ref{eq:HK}), called Kohn--Sham density-functional theory. $G[n]$ is first partitioned into the sum of two other universal density functionals:
\begin{equation}
    \label{eq:KS}
    G[n] = T_s[n] + E_\text{xc}[n]
\end{equation}
the kinetic energy of a system of noninteracting electrons $T_s[n]$, and (now, by definition) the exchange and correlation energy (correction) of the interacting system $E_\text{xc}[n]$. $T_s[n]$ can be calculated exactly, by introducing a set of $N$ one-electron, orthonormal wavefunctions $\{\phi_i\}_{i=1}^N$:
\begin{equation}
    \label{eq:KS2}
    T_s[n] = \sum_{i=1}^N -\frac{1}{2} \int d\rvec ~ \phi_i^*(\rvec) \nabla^2 \phi_i(\rvec)
\end{equation}
(in atomic units) with density $n$:
\begin{equation}
    \label{eq:n}
    n(\rvec) = \sum_{i=1}^N | \phi_i(\rvec) |^2 ~~~ , ~~~ n = \int d\rvec ~ n(\rvec) 
\end{equation}
Given that the magnitude of $E_\text{xc}[n]$ is much smaller than than $T_s[n]$, even approximations to it have resulted in this method becoming one of the most popular for studying the ground-state properties of many-electron systems \cite{RevModPhys.87.897}.

While the above method has had remarkable successes, its applicability is limited by the treatment of the density functionals that appear in Eq.\ (\ref{eq:KS}) \cite{:/content/aip/journal/jcp/136/15/10.1063/1.4704546}. In particular, its accuracy relies (entirely) on the approximation to $E_\text{xc}[n]$. Such is discussed below. And even though $T_s[n]$ can be calculated exactly, it is this which constitutes the major fraction of computational cost. Orthogonalization of $\{\phi_i\}_{i=1}^N$ makes the method scale as $\bigO(N^3)$; and for condensed matter, the need to sample it over the Brillouin zone can add several orders of magnitude in computational cost. 

For the reasons outlined above, there has been considerable effort toward the development of a better approximations to $E_\text{xc}[n]$, as well as orbital-free approximations to $T_s[n]$ \cite{Karasiev20122519} (the latter would avoid the need to introduce $\{\phi_i\}_{i=1}^N$). Consider $E_\text{xc}[n]$, for example. Originally \cite{PhysRev.140.A1133}, a local density approximation was made. The accuracy of this, which is much higher than \textit{a priori} expected, and its computational simplicity have made it possible to accurately model many systems, such as their ground-state energies and structural properties. Still, many other applications require an accuracy at least an order of magnitude better. Improvements have traditionally been based on either approximations derived from quantum mechanics (e.g., Ref.\ \onlinecite{PhysRevLett.77.3865}), or empirical ones containing parameters fit to improve the accuracy on particular chemical systems (e.g., Refs.\ \onlinecite{PhysRevA.38.3098} and \onlinecite{PhysRevB.37.785}). While many of those derived work surprisingly well, they are unable to consistently provide the high accuracy needed for many problems. Recently, however, a different approach was proposed in Ref.\ \onlinecite{PhysRevLett.108.253002}, based on (conventional) machine learning. Unlike traditional approaches, machine learning methods are not based on an assumption as to the underlying model, but rather on the discovery of patterns in high-dimensional data. They therefore provide a powerful and flexible approach to density-functional approximation.

Conventional machine learning methods, however, are very limited in their ability to process raw data in their natural form. Consider linear classification models. These can only divide their input space into half-spaces separated by a hyperplane \cite{Duda_Hart}. They therefore perform poorly on problems where the (input--output) function must be insensitive to irrelevant variations in the input data, such as translations or rotations, while at the same time be very sensitive to small variations in it (which is the case for densities and density functionals \cite{:/content/aip/journal/jcp/139/22/10.1063/1.4834075}). While the invariance problem can be solved by preprocessing the data using good feature extractors, this requires considerable domain expertise. The sensitivity can be improved using generic, nonlinear features, such as kernel methods \cite{Learning_with_Kernels} (e.g., as done in Ref.\ \onlinecite{PhysRevLett.108.253002}, and further studied in Ref.\ \onlinecite{QUA:QUA25040}). However, machine learning algorithms that rely solely on a smoothness prior, with a similarity between examples expressed by a local kernel, are sensitive to the variability of the target \cite{Bengio06thecurse}; in other words, they cannot generalize, and require a number of training cases proportional to the number of variations of the target function. 

In this work, an alternative approach to density-functional approximation is presented, based on deep learning \cite{Nature.521.436}. Deep learning allows computational models that are capable of discovering intricate structure in large and/or high-dimensional data sets, with multiple levels of abstraction. Such methods operate well on raw data (and potentially unlabeled) in their natural form, with the intent that the abstractions make it easier to separate from each (and even extract \cite{Erhan-vis-techreport-2010}) the underlying explanatory factors (features). This disentanglement leads to features in higher layers that are more invariant to some factors of variation (compared to prior layers, including the raw input) and more sensitive to others \cite{NIPS2009_3790}. Importantly, this occurs without the need to introduce feature extractors and/or nonlinear features. In addition, this can resolve nonlocal correlations in the input data, resulting in features that are locally similar between examples.

This Article is organized as follows. Section \ref{sec:methods} discusses the methods, including the development of the deep learning model; Section \ref{sec:results} presents results from the application of this model to the approximation of $T_s[n]$; Section \ref{sec:model_analysis} studies the model in detail; and Section \ref{sec:discussion} discusses the advantages of this model, its extension to any other property that may depend on the density (e.g., $E_\text{xc}[n]$), and other aspects of it, and concludes. The Appendices provide more precise details of the methods. A Supplementary Information (SI) accompanies this Article that provides the raw data used to calculate the results, as well as additional details, discussed in context below. Note that throughout this Article, focus is placed primarily on the deep learning model, and in sufficiently general terms so that it can be extended to other problems.


\section{Methods}
\label{sec:methods}

\subsection{Deep learning model}
\label{sec:results:ML_alg}

The deep learning model developed is based on a generative deep architecture \cite{SIP:9155271}, capable of learning representations of data in terms of separated, high-order features. A generative model makes use of hidden (latent) variables to describe the probability distribution over (visible) data values, by specifying a joint probability distribution over both. The hidden variables introduce correlations between the visible data, and they usually have a simple distribution.

\begin{figure*}
    \centering
    \subfloat[Restricted Boltzmann machine (RBM)]{{\includegraphics[width=8.1cm]{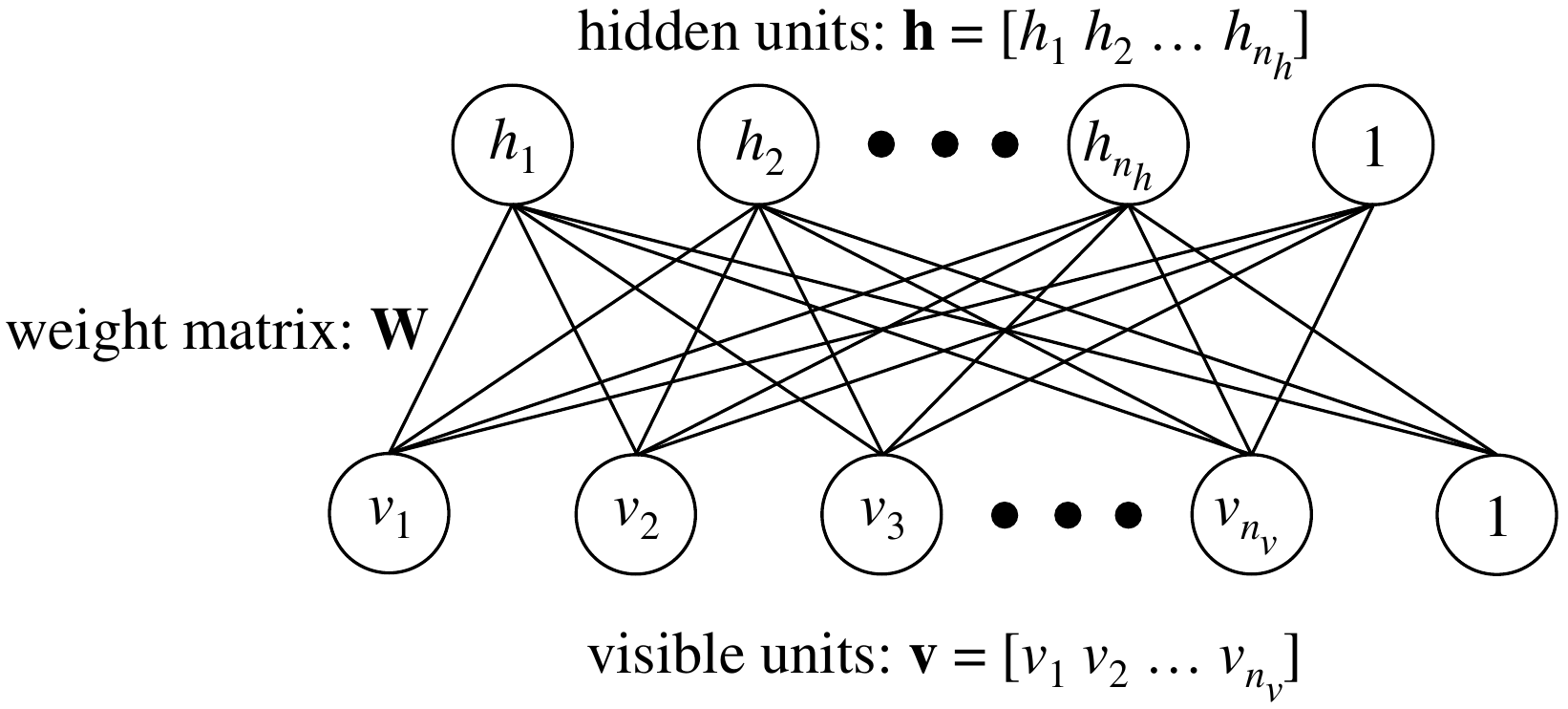}}}
    \qquad
    \qquad
    \subfloat[Deep belief network (DBN)]{{\includegraphics[width=6.75cm]{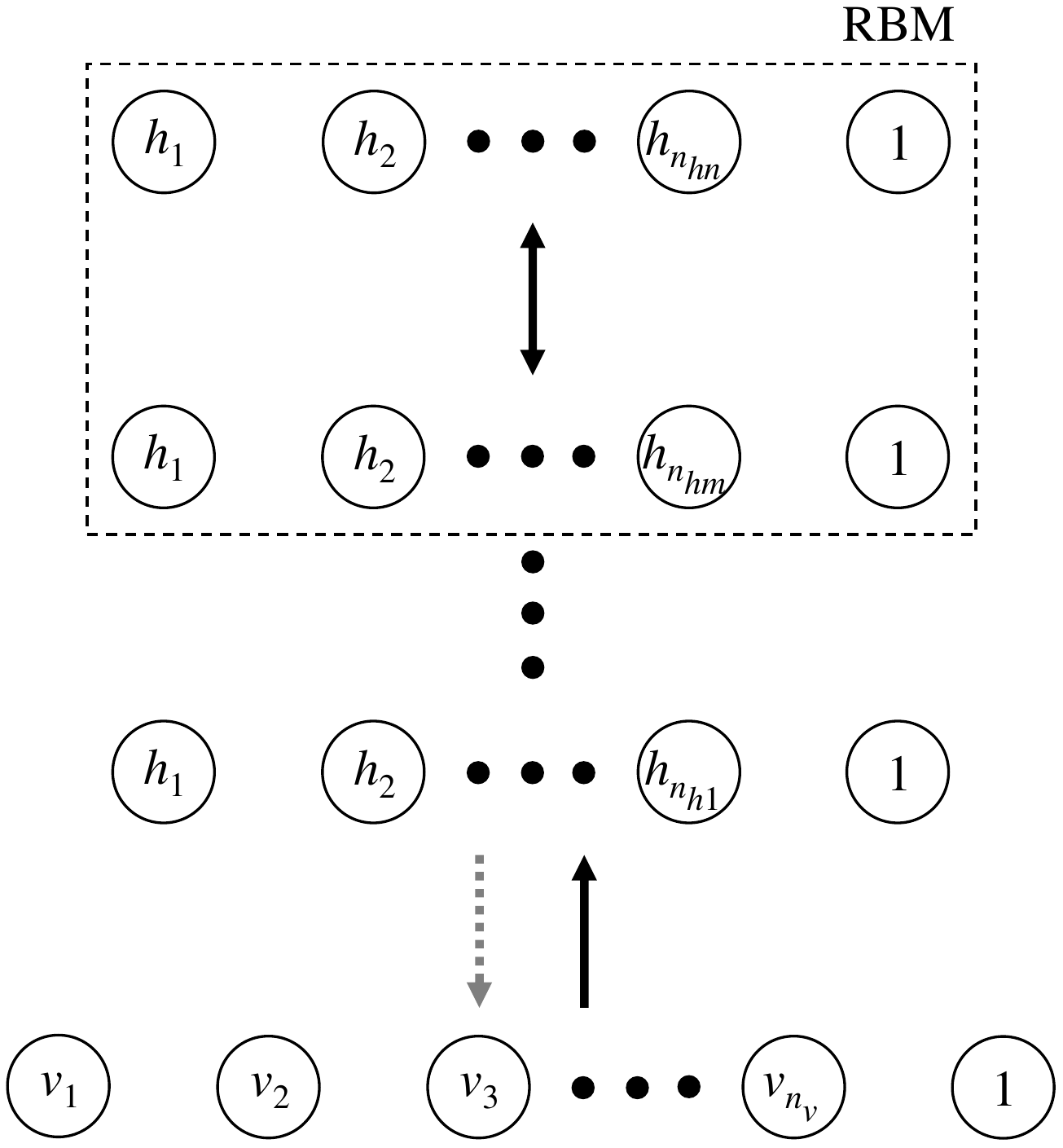}}}
    \qquad
    \qquad
    \subfloat[DBN + Gaussian process (DBN+GP)]{{\includegraphics[width=5.4cm]{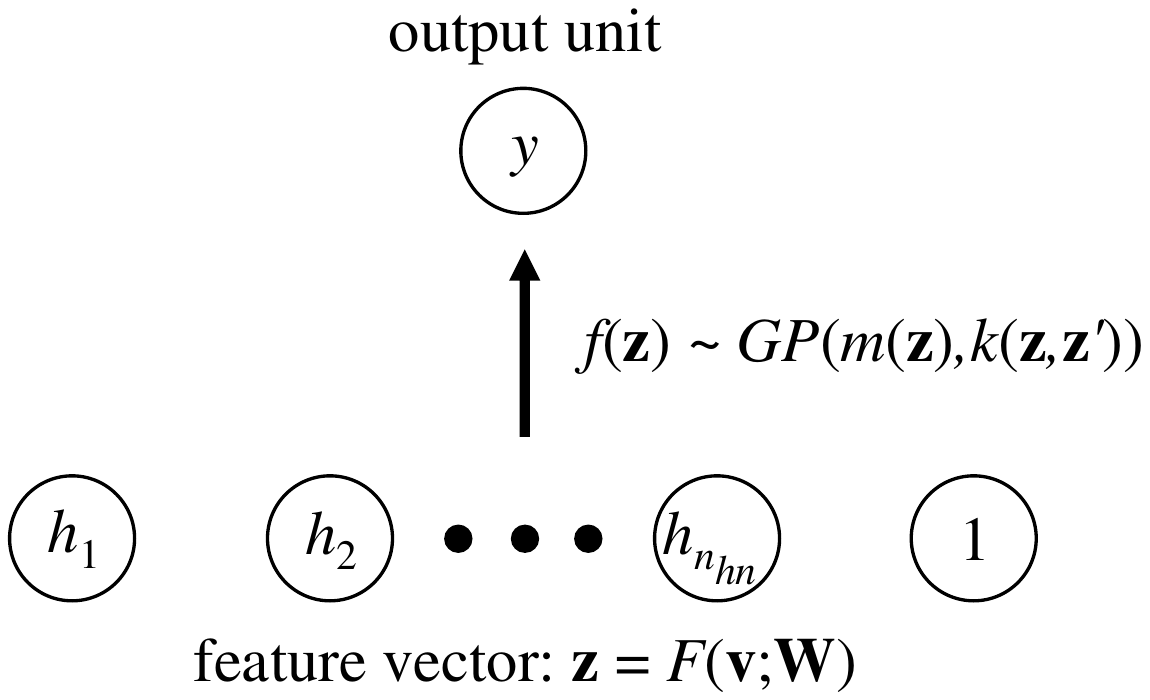}}}
    \caption{Schematic diagram of the deep learning model. Each component is described in detail in the text.}
    \label{fig:DBNGP}
\end{figure*}

The process by which features are learned can understood by considering a restricted Boltzmann machine (RBM), as shown in Fig.\ \ref{fig:DBNGP}(a). 
%
%
An RBM is a two-layer network, where $n_v$ (stochastic) visible units $\mathbf{v} = (v_1, v_2, \dots, v_{n_v})$ are connected to $n_{h}$ stochastic hidden ones $\mathbf{h} = (h_1, h_2, \dots, h_{n_{h}})$, using symmetrically-weighted connections $\mathbf{W}$. Training consists of adjusting the latter to maximize the product of marginal probabilities of $\mathbf{v}$, $p(\mathbf{v})$ \footnote{In this work, the input data $\mathbf{v}$ is continuous. Therefore, $p(\mathbf{v})$ corresponds to a probability density function. The notation was chosen to be consistent with that commonly used.}, assigned to a set of $M$ (possibly unlabeled) input data points $\mathbf{V} = \{ \mathbf{v}_i \}_{i=1}^{M}$:
\begin{equation}
    \label{eq:RBM_Pv}
    \arg \max_\mathbf{W} \prod_{\mathbf{v} \in \mathbf{V}} p(\mathbf{v})
\end{equation}
where:
\begin{equation}
    \label{eq:Pv}
    p(\mathbf{v}) = \frac{1}{Z} \sum_\mathbf{h} e^{-E(\mathbf{v}, \mathbf{h})}
\end{equation}
is defined in analogy with the Boltzmann distribution \cite{mcquarrie2000statistical}, and is obtained from a sum over all of the possible hidden unit configurations, where $E(\mathbf{v}, \mathbf{h})$ is the ``energy'' of a joint configuration $(\mathbf{v}, \mathbf{h})$ \cite{Hopfield01041982}:
\begin{equation}
    \label{eq:E_vh}
    E(\mathbf{v}, \mathbf{h}) = - \mathbf{a}^\mathrm{T} \mathbf{v} - \mathbf{b}^\mathrm{T} \mathbf{h} - \mathbf{v}^\mathrm{T} \mathbf{W} \mathbf{h}
\end{equation}
where $\mathbf{a}$ and $\mathbf{b}$ are vectors of bias weights for the visible and hidden units, respectively (bias units are denoted by $1$s in Fig.\ \ref{fig:DBNGP}), and $Z$ is a partition function defined as a sum over all possible configurations:
\begin{equation}
    \label{eq:Z}
    Z = \sum_{\mathbf{v}, \mathbf{h}} e^{-E(\mathbf{v}, \mathbf{h})}
\end{equation}
Equation \ref{eq:E_vh} applies to binary visible and hidden units; for continuous ones, addition of the following terms:
\begin{equation}
    \label{eq:E_vh_cont}
    \int_{a_v}^{b_v} d\mathbf{v} ~ f_v(\mathbf{v}) ~~~ , ~~~ \int_{a_h}^{b_h} d\mathbf{h} ~ f_h(\mathbf{h}) 
\end{equation}
where $f_{v(h)}$ are the visible (hidden) unit activation functions, ensures that Eq.\ (\ref{eq:Pv}) can be normalized. After maximizing Eq.\ (\ref{eq:RBM_Pv}), an RBM provides a closed-form representation of $p(\mathbf{v})$, denoted herein as $p(\mathbf{v}; \mathbf{W})$.

Often, the representational power of a single RBM is limited, or it is difficult to separate the input data into simple distributions. In this case, RBMs can be stacked, learning successive layers of abstractions. The resulting model is called a deep belief network (DBN), as shown in Fig.\ \ref{fig:DBNGP}(b) (note the indication of layer indices $l$ on $n_{hl}$), which can be extremely powerful. In addition to the increased representational power, DBNs offer many other attractive features; a comprehensive discussion of these can be found in Ref.\ \onlinecite{SIP:9155271}.

Following training, the DBN is used to initialize a nonlinear mapping:
\begin{equation}
    \label{eq:F}
    F\colon \mathbf{V} \mapsto \mathbf{Z}
\end{equation}
parameterized by the weights $\mathbf{W}$ of the DBN, which maps the input vector space $\mathbf{V}$ to its feature space $\mathbf{Z}$. Elements of the latter corresponds to the high-level features of the DBN, following a deterministic forward-propagation of the former. Note that $F$ (in fact, the entire model --- see the further discussion in Appendix \ref{sec:methods:ML_alg}) is initialized in an entirely unsupervised way.

Finally, with $F$ specified, supervised learning is used to find a mapping $f$ from the high-level features $\mathbf{z}$ to an output $y$. In this work, it is assumed that the $y$ differs from the function(al) $f$ (herein, $f$ is technically a functional) by additive noise $\varepsilon$ that follows an independent and identically distributed Gaussian distribution with zero mean and variance $\sigma_n^2$:
\begin{equation}
    \label{eq:prob_regression}
    y = f(\mathbf{z}) + \varepsilon ~~~ , ~~~ \varepsilon \sim \mathcal{N}(0,\sigma_n^2)
\end{equation}
That is, we assume that $f$ is distributed according to a Gaussian process (GP), with mean function $m(\mathbf{z})$ (herein chosen to be $0$) and covariance function $k(\mathbf{z},\mathbf{z}')$. Realize that any other supervised method (e.g., a multilayer perceptron) could be used to find $f$; such is discussed in the SI. This is because the invariances and sensitivity are modeled via the DBN. The use of a GP should therefore be understood as a choice made without loss of generality. Because of this, details of the GP are deferred until Section \ref{sec:methods:ML_alg}. With this choice though, the complete deep learning model is referred to as the DBN+GP model, as shown in Fig.\ \ref{fig:DBNGP}(c).

\subsection{Model system}
\label{sec:results:model}

The model system considered herein is analogous to Ref.\ \onlinecite{PhysRevLett.108.253002}: $N$ noninteracting, spinless electrons confined to a 1D box, with hard walls and a continuous potential. Solving this model for $n$ and $T_s[n]$ for randomly generated potentials provides the data set for which to train and test the deep learning model. Appendix \ref{sec:methods:model} provides a more thorough discussion, with specific details.

\subsection{Performance evaluation}
\label{sec:results:performance_measures}

Following training, the performance of the DBN+GP model in approximating $T_s[n]$ (referred to below as simply ``performance'') was assessed by testing it on unseen data. Performance statistics were selected so as to give a comprehensive assessment of a given model, as well as allow a direct comparison between different ones: the normalized mean squared error (NMSE) \cite{Hanna_NMSE_1985}, which describes the amount of relative scatter, and tends not to be biased toward models that under- or overpredict; the normalized mean bias factor (NMBF) \cite{ASL:ASL125}, a symmetric measure that describes the amount of bias present; and the square of the sample correlation coefficient ($r^2$) \cite{Pearson01011895}, which describes the proportion of variance in the input data that is accounted for. Formulas, a discussion of their error estimates, and additional details are given in Appendix \ref{sec:methods:statistics}.

\subsection{Computational Details}
\label{sec:methods:comp_details}

Details of the DBN+GP model and its training that are relevant for the following discussion are as follows, unless otherwise specified. For the DBN: two RBMs were stacked, with the number of hidden units in layers $1$ and $2$ $n_{h1}$-$n_{h2} = 50$-$25$; $\mathbf{V}$ consisted of a set of $M_\text{ul} = 500$ (unlabeled) density vectors $\bm{n}$ (see Appendix \ref{sec:methods:model}). For the GP: the input consisted of $M_\text{l} = 50$ (labeled) training vectors (randomly selected from $M_\text{ul}$) containing the value of the ($n_{h2}$) top-level features mapped to by $F$;  the output was taken to be $y = T_s[n]$. While these choices were made primarily for demonstrative purposes, they are nonetheless justified further below. Additional details are given in Appendices \ref{sec:methods:ML_alg} and \ref{sec:methods:model}


\section{Results}
\label{sec:results}

\subsection{Kinetic-energy density functional}
\label{sec:results:KEDF}

A useful density functional must be accurate over a range of densities. Table \ref{Tb:n} shows the performance as $N$ is increased from $2$ to $8$. 
\begin{table}
    \caption{Performance for different numbers of electrons $N$, for $N = 2$ to $8$.}
    \label{Tb:n}
    \begin{ruledtabular}
        \begin{tabular}{e{1.0} d{1.8} d{2.6} d{1.6}}
            \multicolumn{1}{c}{$N$}                        &  \multicolumn{1}{c}{NMSE ($\times 10^{-6}$)}      &  \multicolumn{1}{c}{NMBF ($\times 10^{-4}$)}     &  \multicolumn{1}{c}{$r^2$}      \\
            \hline
            \hline
            2.                              &  3.1(7)                     &  -1.6(6)                      &  0.977(4)      \\
            \hline
            3.                              &  0.34(7)                  &  -1.0(2)                       &  0.93(1)         \\
            \hline
            4.                              &  0.035(5)                &  -0.06(6)                    &  0.960(5)      \\
            \hline
            5.                              &  0.0076(8)              &  0.15(3)                     &  0.951(5)     \\      
            \hline
            6.                              &  0.0017(3)              &  -0.07(1)                    &  0.959(5)      \\
            \hline
            7.                              &  0.0007(1)              &  0.002(8)                   &  0.948(7)         \\
            \hline
            8.                              &  0.00015(2)            &  -0.015(4)                 &  0.970(3)     
        \end{tabular}
    \end{ruledtabular}
\end{table}
It can be seen in the NMSE that the relative scatter by the model is very small, on the order of $10^{-6}$ for small $N$, decreasing to $10^{-10}$ as $N$ increases. Similar behavior is seen in the NMBF, which shows that there is very little bias, and which decreases in magnitude from $10^{-4}$ to $10^{-6}$. These trends can be understood by considering that as $N$ increases, the density becomes more uniform \cite{PhysRevLett.100.256406}; so predictions are made for smaller changes of $T_s[n]$ relative to larger magnitudes. Therefore, the NMSE and/or NMBF values in Table \ref{Tb:n} should not be interpreted as the model performing better for large $N$. A better comparative measure (in this case) is $r^2$, which shows that the model is able to account for approximately $96\%$ of the variance in the input data, independent of $N$. 

Because the results in Table \ref{Tb:n} indicate that the performance is (relatively) independent of $N$, only $N = 4$ is considered below. This choice provides a good balance between the total variation in the input data and its uniformity.

\subsection{Self-consistent densities}
\label{sec:results:SCn}

In practice, not only is an accurate approximation to $T_s[n]$ needed, but also the ability to use it to find self-consistent densities. Table \ref{Tb:SCn} shows the performance when self-consistent densities, the latter obtained using the approach outlined in Appendix \ref{sec:methods:SCn}.
\begin{table}
    \caption{Performance, using self-consistent densities.}
    \label{Tb:SCn}
    \begin{ruledtabular}
        \begin{tabular}{c c c}
            NMSE ($\times 10^{-6}$) & NMBF ($\times 10^{-4}$) & $r^2$ \\
            \hline
            \hline
            0.46(3)                     &  -4.0(2)                      &  0.81(1)    
        \end{tabular}
    \end{ruledtabular}
\end{table}
Note that the same model used to calculate the results in Table \ref{Tb:n} (for $N = 4$) was used, allowing a direct comparison between the two tables. The DBN+GP model is seen to retain most of its predictive ability. The differences can be understood by considering the NMBF. While its magnitude remains low (on the order of $10^{-3}$ to $10^{-4}$), it decreases by about an order of magnitude (in an absolute sense). This can be attributed to that the accuracy of $F$ is limited by the representation $p(\mathbf{v};\mathbf{W}) \approx p(\mathbf{v})$; this affects the ability of the supervised learning algorithm to correctly represent the function(al) $f$, whose accuracy might also be limited. Without the complete and accurate $F$ and $f$, the DBN+GP model (probably) has sufficient flexibility to find a density for which it can provide a underestimation of $T_s[n]$ (to minimize the energy). The increase in NMSE and decrease in $r^2$ suggest that the extent of this is variable, since these measures are (relatively) independent of bias.

Discussions about $F$ and $f$, including universal approximation properties, are provided in Sections \ref{sec:results:F} and \ref{sec:results:fz}, respectively. Improvements to the approach for finding self-consistent densities are discussed in Appendix \ref{sec:methods:SCn}.


\section{Model Analysis}
\label{sec:model_analysis}

\subsection{The mapping $\bm F$}
\label{sec:results:F}



The representational power of $F$ is determined by that of the DBN used to initialize it. In the case of binary inputs (the continuous extension is considered below), it has recently been proven \cite{LeRoux:2008:RPR:1374176.1374187} that adding hidden units (to an RBM) strictly improves modeling power. The results in Table \ref{Tb:architecture} show that the DBN+GP model is consistent with this; the performance improves with increasing the number of hidden units.
%
%
\begin{table}
    \caption{Improvement in performance as the representational power of $F$ is increased, by increasing the number of hidden units $n_{h1}$ and $n_{h2}$ of the DBN. A reference point, identical between Tables \ref{Tb:architecture}--\ref{Tb:labled_data}, is marked by a $^*$.}
    \label{Tb:architecture}
    \begin{ruledtabular}
        \begin{tabular}{e{4.3} d{1.6} d{2.5} d{1.6}}
            \multicolumn{1}{c}{$n_{h1}$-$n_{h2}$}        &  \multicolumn{1}{c}{NMSE ($\times 10^{-6}$)}      &  \multicolumn{1}{c}{NMBF ($\times 10^{-4}$)}     &  \multicolumn{1}{c}{$r^2$}      \\
            \hline
            \hline
            25-.10                          &  0.13(2)                  &  -0.3(2)                       &  0.87(2)        \\
            \hline
            25-.25                          &  0.059(7)                &  -0.4(1)                      &  0.932(8)      \\
            \hline
            50-.25^*                      &  0.034(3)                &  -0.2(1)                       &  0.962(3)         \\
            \hline
            125-.50                        &  0.020(3)                &  -0.17(5)                    &  0.976(3)     
        \end{tabular}
    \end{ruledtabular}
\end{table}
It also appears that the (relative) uncertainties of the performance statistics decrease with increasing architecture size (other factors remaining fixed); however, additional results would be needed to confirm this trend. It has also been proven \cite{LeRoux:2008:RPR:1374176.1374187} that the improvement in representational power by adding a second layer in a DBN is limited by that of the first layer. This is also reflected in Table \ref{Tb:architecture}; compare, for example, $n_{h1}$-$n_{h2}$ = $25$-$25$ to $50$-$25$. Consideration of these results, balanced by the complexity of the DBN, suggests $n_{h1}$-$n_{h2}$ = $50$-$25$ as a reasonable architecture for demonstrative purposes.

The proofs (mentioned above, in Ref.\ \onlinecite{LeRoux:2008:RPR:1374176.1374187}) can be extended to address the theoretical limits of the representational power of $F$. Recently, universal approximation properties for DBNs with continuous visible units and binary hidden units have been proven \cite{ICML2013_krause13}; in particular, one theorem proves the existence of a DBN with a finite number of hidden units that can approximate $p(\mathbf{v})$ (for any $p(\mathbf{v})$) arbitrarily well. This implies the existence of a universal mapping $F$, with a finite number of features, that is not bound in representational power.


%
%
For a fixed DBN architecture, the resolution of $F$ is a function of the data used to train it. Equation (\ref{eq:RBM_Pv}) shows that an RBM can only learn the distribution of the training data, which might be an incomplete representation of $p(\mathbf{v})$. The effect of this is illustrated in Table \ref{Tb:unlabled_data}. 
\begin{table}
    \caption{Improvement in performance as the resolution of $F$ is increased, by increasing the number of (unlabeled) training points $M_\text{ul}$ used to train the DBN. A reference point, identical between Tables \ref{Tb:architecture}--\ref{Tb:labled_data}, is marked by a $^*$.}
    \label{Tb:unlabled_data}
    \begin{ruledtabular}
        \begin{tabular}{e{4.1} d{1.6} d{2.5} d{1.6}}
            \multicolumn{1}{c}{$M_\text{ul}$}        &  \multicolumn{1}{c}{NMSE ($\times 10^{-6}$)}      &  \multicolumn{1}{c}{NMBF ($\times 10^{-4}$)}     &  \multicolumn{1}{c}{$r^2$}      \\
            \hline
            \hline
            100.                         &  0.046(4)                &  -0.37(6)                    &  0.948(4)        \\
            \hline
            200.                         &  0.043(5)                &  -0.23(7)                    &  0.950(6)      \\
            \hline
            500.^*                     &  0.034(3)                &  -0.2(1)                       &  0.962(3)         \\
            \hline
            1000.                       &  0.028(3)                &  -0.24(7)                    &  0.970(3)     
        \end{tabular}
    \end{ruledtabular}
\end{table}
Increasing $M_\text{ul}$ from $100$ to $1000$, for example, decreases the NMSE by $\jmmapprox 40\%$ and increases $r^2$ by $2\%$; there is no conclusive trend in the NMBF. Realize that these results are independent of labeled training data, which is kept fixed (at $M_\text{l} = 50$). These results should not be taken to imply that large data sets are required to achieve high accuracy; even a low amount of points is relatively accurate. This is supported by the recent findings \cite{AISTATS2011_BengioBBBBCCCEEGMLPRSS11} that deep learners benefit from out-of-distribution examples; in the present context, this is analogous to an incomplete sampling from some probability distribution.

The SI discusses the issues presented in this section in more detail, including the relationship between DBN architecture, $M_\text{ul}$, and the generalization capability of the resulting DBN+GP model, as well as justifies the choice of $M_\text{ul} = 500$.

\subsection{The function(al) $\bm f$}
\label{sec:results:fz}




Insight into the DBN+GP model and its performance can be obtained by looking at the efficiency of $f$ to map the high-level features of $F$ to a desired output. The efficiency $\eta$ of $f$ can be defined as (for its derivation, see the SI):
%
\begin{equation}
    \label{eq:efficiency}
    \eta = \left( \frac{\text{ACC}}{M_\text{l}} G[V(y, \Omega)] \right) \bigg / \eta_0
\end{equation}
where $\text{ACC}$ is the accuracy of the model, $G[V(y, \Omega)]$ is an (unknown) functional of the total variation of the target function $y$ $V(y, \Omega)$, defined on a bounded open set $\Omega \subseteq \mathbb{R}^d$, where $d$ is the dimension of the underlying manifold which the data lie, describing the ``complexity'' of a target function, $M_\text{l}$ has been defined previously, and $\eta_0$ is a normalization factor.

\begin{figure*}
    \includegraphics[scale = 0.3]{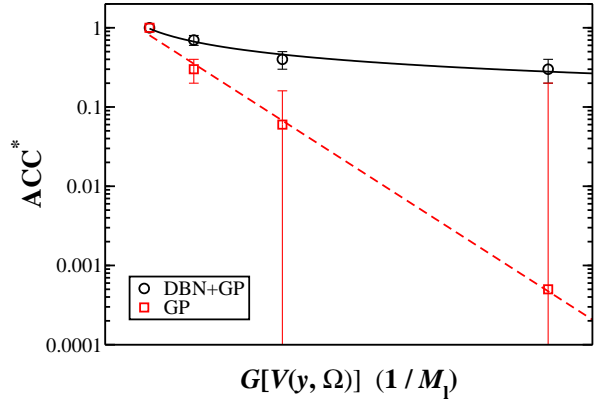}
    \caption{Normalized accuracies ($\mathrm{ACC}^*$) of the DBN+GP model in comparison to a GP, as a function of target variability. Note that the error bars appear exaggerated on the log scale.}
    \label{fig:efficiency}
\end{figure*}

The efficiency of a model on a given problem gives a direct indication of its applicability to those more (or less) complex; this can be seen in Eq.\ (\ref{eq:efficiency}). Figure \ref{fig:efficiency} shows this for the results in Table \ref{Tb:labled_data}, in comparison to using a GP directly. 
\begin{table}
    \caption{Improvement in performance as the accuracy of $f$ is improved, by increasing the number of (labeled) training points $M_\text{l}$ used to train the GP. A reference point, identical between Tables \ref{Tb:architecture}--\ref{Tb:labled_data}, is marked by a $^*$.}
    \label{Tb:labled_data}
    \begin{ruledtabular}
        \begin{tabular}{e{3.1} d{1.6} d{2.5} d{1.6}}
           \multicolumn{1}{c}{$M_\text{l}$}     &  \multicolumn{1}{c}{NMSE ($\times 10^{-6}$)}      &  \multicolumn{1}{c}{NMBF ($\times 10^{-4}$)}     &  \multicolumn{1}{c}{$r^2$}      \\
            \hline
            \hline
            20.                         &  0.044(3)                &  -0.5(1)                        &  0.951(3)        \\
            \hline
            50.^*                      &  0.034(3)                &  -0.2(1)                       &  0.962(3)         \\
            \hline
            100.                       &  0.020(2)                &  -0.16(4)                     &  0.975(3)         \\
            \hline
            200.                       &  0.014(1)                &  -0.10(2)                     &  0.983(2)     
        \end{tabular}
    \end{ruledtabular}
\end{table}
Note that an increase in $G[V(y, \Omega)]$ (for a fixed $M_\text{l}$) is equivalent to that in $1 / M_\text{l}$ (for a fixed $G[V(y, \Omega)]$), assuming that $\eta_0$ can be calculated the same. The DBN+GP model is seen to be much more efficient, and robust against an increase in variability (for a fixed $M_\text{l}$); the GP shows an exponential decrease in accuracy. These results are consistent with the theoretical arguments of Ref.\ \onlinecite{Bengio06thecurse}, that local kernels are sensitive to $V(y, \Omega)$, while there exist nonlocal learning algorithms that are not (or at least have the potential to learn about functions with a high $V(y, \Omega)$, without requiring a proportional number of training examples or using very specific prior domain knowledge); and also the results of Ref.\ \onlinecite{NIPS2009_3790}, that more abstract features in higher layers of deep architectures are more robust to unanticipated sources of variance. The increase in efficiency can be quite important, when considering that $V(y, \Omega)$ may exponentially increase with $d$.

The high efficiency of the DBN+GP model suggests that for the simple problem defined in Section \ref{sec:results:model} and Appendix \ref{sec:methods:model} (which has a low $G[V(y, \Omega)]$) that it should still be relatively accurate for a low $M_\text{l}$. This is confirmed in Table \ref{Tb:labled_data}, where accurate results (e.g., NMSE $\approx 10^{-7}$ to $10^{-8}$ and $r^2 \approx 0.95$) are shown with even as few as $M_\text{l} = 20$ data points ($\jmmapprox 1$ per feature). Though, the accuracy is expected to increase with $M_\text{l}$; indeed, Table \ref{Tb:labled_data} shows that the NMSE decreases to $10^{-8}$ and $r^2$ increases to over $0.98$ for $M_\text{l} = 200$. For demonstrative purposes, $M_\text{l} = 50$ provides a reasonable level of accuracy.

\subsection{Generative sampling}
\label{sec:results:gen_sampling}



\begin{figure*}
    \centering
    \subfloat[$p(\mathbf{v};\mathbf{W})$]{{\includegraphics[width=7.5cm]{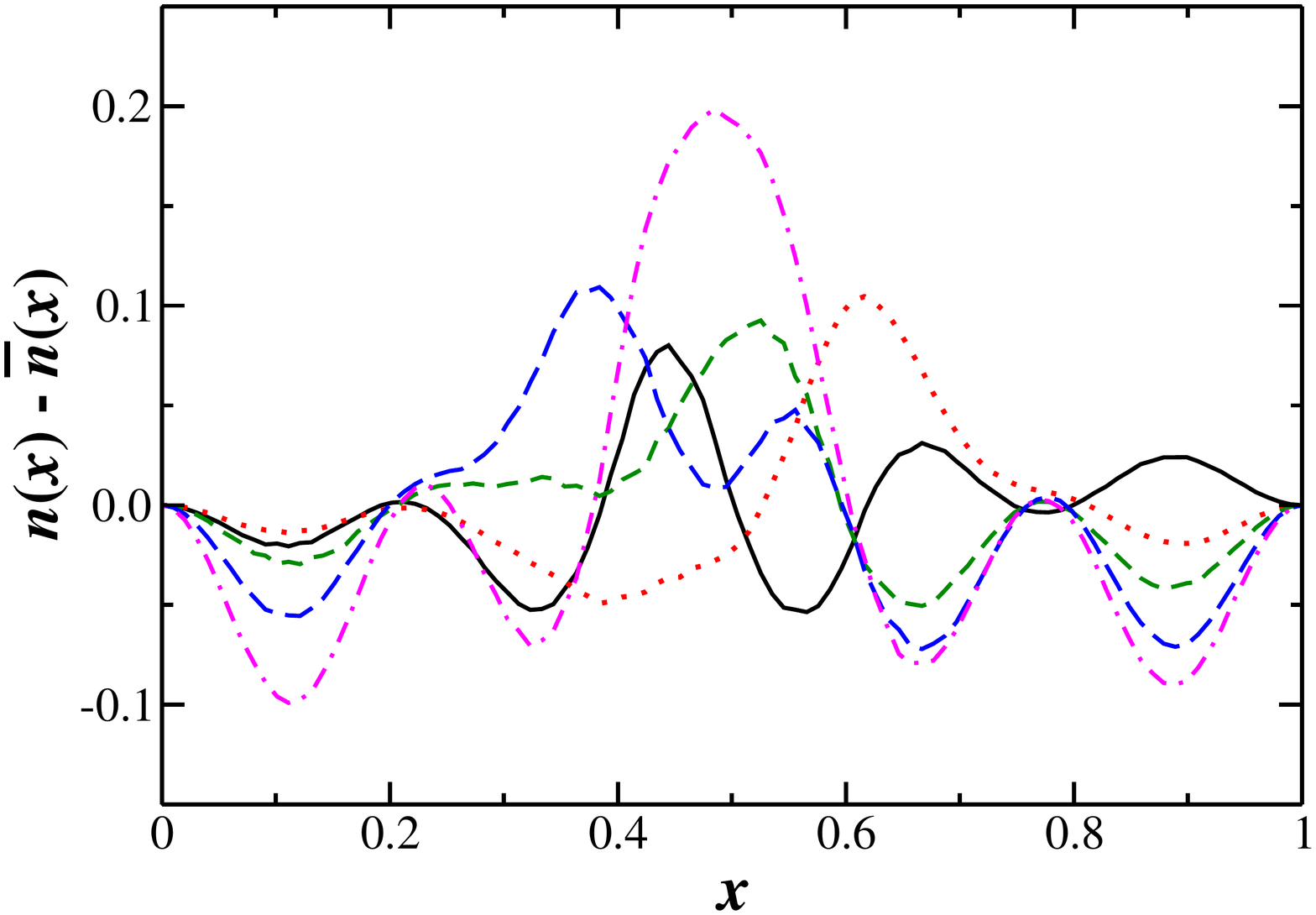}}}
    \qquad
    \subfloat[$p(\mathbf{v})$]{{\includegraphics[width=7.5cm]{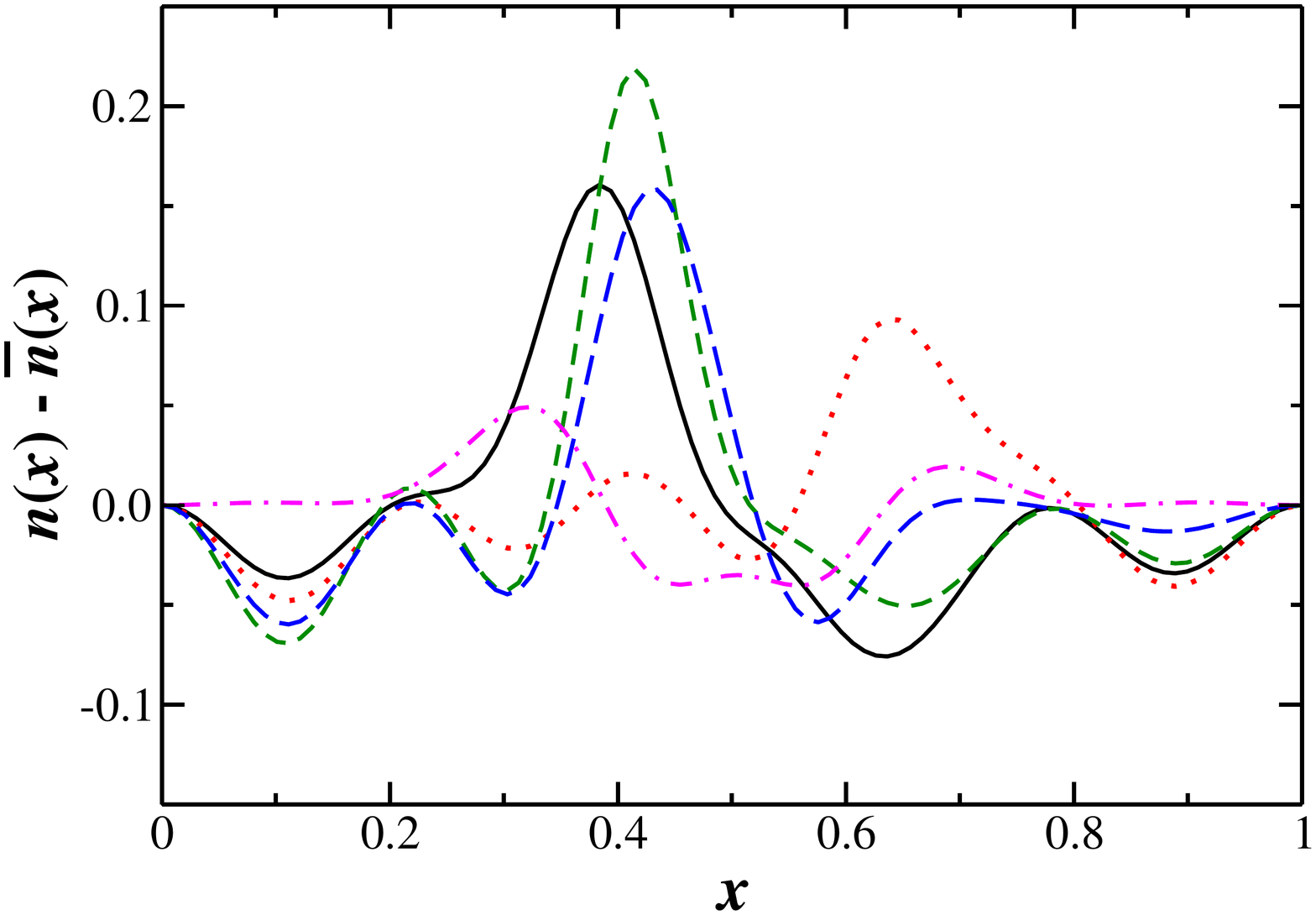}}}
    \caption{Samples of $\mathbf{v}$ ($\mathbf{n}$), shifted to $\mathbf{v} - \mean{\mathbf{v}}$, drawn from (a) $p(\mathbf{v};\mathbf{W})$ and (b) $p(\mathbf{v})$. Note that a direct comparison between the individual curves in (a) and (b) cannot be made.}
    \label{fig:gen_samping}
\end{figure*}

Based on Eq.\ (\ref{eq:RBM_Pv}), each layer of features produced by the mapping $F$ must allow for the most probable reconstruction of those prior. Therefore, they must contain most of the same information, but expressed in a way that makes explicit the higher-order structure. An indirect way to see this is compare samples drawn from $p(\mathbf{v};\mathbf{W})$ to those from $p(\mathbf{v})$. The former can be obtained via generative sampling, the precise details of which are discussed in Appendix \ref{sec:methods:gen_sampling}. For the model system considered herein (Section \ref{sec:results:model} and Appendix \ref{sec:methods:model}), this comparison is easy to make qualitatively, since samples of $\mathbf{v}$ have a small deviation about $\mean{\mathbf{v}}$ (without translations or rotations). This is shown in Fig.\ \ref{fig:gen_samping}.
%
Note that the same model used to calculate results in Tables \ref{Tb:n} (for $N = 4$) and \ref{Tb:SCn} was used here. It can be seen that samples drawn from $p(\mathbf{v};\mathbf{W})$ and $p(\mathbf{v})$ are qualitatively very similar.




\section{Discussion and Conclusion}
\label{sec:discussion}




A computational model based on deep learning \cite{Nature.521.436}, the DBN+GP model, was developed and applied to the problem of density-functional approximation. Through a detailed analysis in Section \ref{sec:results}, this model was shown to perform well on approximating $T_s[n]$ for noninteracting electrons in a 1D box. There is even room for improvement, by optimizing its parameters. In addition to performance, it was shown (both directly and indirectly) to offer several advantages over conventional machine learning. Perhaps most importantly, it was initialized directly from the input data, in an entirely unsupervised way and without introducing feature extractors and/or nonlinear features. 


Even though the DBN+GP model was applied to approximate $T_s[n]$, its extension to any other property which may depend on the density is straightforward. This is because the features underlying the input data are (obviously) independent of the output. The mapping between them is secondary; output information is used determine relationships described by this mapping and perhaps only to refine the features. This can be advantageous, because the model can make efficient use of possibly very large data sets to learn its disentangled high-level features, without requiring it to be labeled. It is therefore particularly well suited for approximating properties for which calculating representative input data is inexpensive, while (accurate) labeled data is not. This is the case for densities and density functionals --- in particular $E_\text{xc}[n]$, which can be calculated \cite{PhysRevLett.45.566,PhysRevLett.87.036401}, but at a high computational cost. 


The developed method also offers an approach by which to obtain physical insight about a system. Since the many-body ground state is a unique functional of the density \cite{PhysRev.136.B864}, insight should be obtainable by extracting information about the learned features. An initial analysis of this was given in Section \ref{sec:results:gen_sampling}; in particular, their collective ability to reconstruct samples from $p(\mathbf{v};\mathbf{W}) \approx p(\mathbf{v})$. An analysis of the individual features may prove interesting though, and several techniques have recently been developed \cite{Erhan-vis-techreport-2010} to provide qualitative interpretations of them, and the invariances that have been learned. Once these are understood, even additional insight may be obtainable by analyzing their mapping to an output(s).


Before concluding, it is important to discuss the extension of this method to actual systems, and the practical issue of computational scaling. Most systems are more complex than that considered herein (for demonstrative purposes). Even though the efficiency analysis in Section \ref{sec:results:fz} demonstrated that the DBN+GP model is not sensitive to the variability of the target, its extension to more complex systems will still require the use of large(r) data sets. To see this: The efficiency advantage is provided by the DBN, and its ability to separate the explanatory factors in the data; but the number of such factors is directly related to the complexity of the system. As discussed in Section \ref{sec:results:F} though, the representational power $F$ is directly related to the architecture of the DBN, and its resolution is determined by $M_\text{ul}$; neither of which is bounded. This means that the method is systematically improvable. Practically important is that network storage and evaluation scales linearly with $n_v$ ($n_h$), for a fixed $n_h$ ($n_v$); and training scales linearly in both time and (storage) space with $M_\text{ul}$. Additional scaling issues are discussed in the SI.


While the DBN+GP model was developed for density-functional approximation, the above advantages are not limited to this problem. In the physical sciences, conventional machine learning methods have shown promise in fields ranging from condensed matter physics \cite{RevCompChem.29.4} to quantum chemistry \cite{QUA:QUA24954}. The developed approach may be useful in these other contexts as well.


\appendix


\section{Deep learning model}
\label{sec:methods:ML_alg}

For each RBM, training proceeded as follows: Initial weights were selected following the approach described in Ref.\ \onlinecite{masters2015deep}; $200$ trials were performed. Satisfying Eq.\ (\ref{eq:RBM_Pv}) was accomplished by the contrastive divergence algorithm \cite{Hinton:2002:TPE:639729.639730}; $5000$ Markov chain Monte Carlo steps were performed, using  learning rate of $0.01$ and a momentum parameter of $0.05$, with the chain length slowly increased from $1$ to $4$. $L_2$ regularization was used to control model complexity; a penalty of $0.0001$ was used. A penalty term was added to encourage sparse activities in the hidden units; a target activation of $0.1$ was set. Finally, a mean-field approximation was made for this visible units (i.e., only the hidden ones were stochastically sampled). Note that the input to the first RBM was normalized, by linearly scaling each dimension (grid point --- see Appendix \ref{sec:methods:model}), using the range of the unlabeled training data.

A DBN was formed by stacking RBMs, using the greedy training algorithm \cite{Hinton:2006:FLA:1161603.1161605}. Note that once an RBM was trained, its weights were fixed (in the DBN). This results in only the top two layers having undirected, symmetric connections; the lower ones receive top-down, directed connections from the layer above; this is indicated in Fig.\ \ref{fig:DBNGP}(b). 

With the mapping $F$ specified completely by the DBN, and according to Eq.\ (\ref{eq:F}), a GP prior was placed over the underlying latent function, so that \textit{a priori} $p(\mathbf{f}|\mathbf{Z}_\text{l}) = \mathcal{N}(\mathbf{f}|m(\mathbf{z}),\mathbf{K}(\mathbf{z},\mathbf{z}'))$, where $\mathbf{Z}_\text{l} = \{ \mathbf{z}_i \}_{i = 1}^{M_\text{l}}$ is the set of features for labeled input data, $\mathbf{f} = [f(\mathbf{z}_1) ~ f(\mathbf{z}_2) ~ f(\mathbf{z}_{M_\text{l}})]^\mathrm{T}$, $\mathbf{K}(\mathbf{z},\mathbf{z}')$ is the covariance matrix with elements $K_{ij} = k(\mathbf{z}_i, \mathbf{z}_j)$, and $m(\mathbf{z})$ has been defined previously. Note that for $m(\mathbf{z}) = 0$, a GP is completely described by $k(\mathbf{z}, \mathbf{z}')$.

The features $\mathbf{z}$ mapped to by $F$ should have a simple distribution, separable from the input data (since the hidden units of an RBM/DBN usually do, as discussed in Section \ref{sec:results:ML_alg}). The prior assumption can therefore be made that if $\mathbf{z}$ and $\mathbf{z}'$ are similar according to some distance measure, their values should be highly correlated. A natural choice is $k(\mathbf{z}, \mathbf{z}') \propto \exp(-\lVert \mathbf{z} - \mathbf{z}' \rVert^2)$; therefore, the spherical Gaussian kernel was used:
\begin{equation}
    \label{eq:Gaussian_kernel}
    k(\mathbf{z}, \mathbf{z}') = \alpha \exp(-\frac{1}{2 \beta} \lVert \mathbf{z} - \mathbf{z}' \rVert^2)
\end{equation}
which is parameterized by $\bm{\theta} = \{\alpha, \beta\}$, a magnitude $\alpha$ and length-scale $\beta$; further, since $\{z | 0 \le z \le 1 \}$ for each element $z$ of $\mathbf{z}$, the assumption can be made that $k(\mathbf{z}, \mathbf{z}')$ has the same $\beta$ for each dimension. As discussed in Section \ref{sec:results:ML_alg}, it is assumed that one does not have access to the values of $f$ themselves, but noisy versions thereof (see Eq.\ (\ref{eq:prob_regression})). The prior on the noisy observations becomes:
\begin{equation}
    \label{eq:prior_on_noise}
    \operatorname{cov}(y_i, y_j) = k(\mathbf{z}_i, \mathbf{z}_j) + \sigma_n^2 \delta_{ij}
\end{equation}

Note that because $k(\mathbf{z}, \mathbf{z}')$ is a function of (only) $F$, it too is initialized in an unsupervised way. The use of a GP as a supervised learning method can therefore be viewed as using a DBN to learn the covariance kernel for a GP \cite{NIPS2007_3211}. In this respect, this approach can be seen as complementary to that in Ref.\ \onlinecite{PhysRevLett.108.253002}; different kernels for which were studied in Ref.\ \onlinecite{QUA:QUA25040}, and found to significantly influence the results. Realize though that the invariance--sensitivity problem (in this context, the kernel $k(\mathbf{z}, \mathbf{z}')$) is solved entirely by the DBN, prior to the use of a GP at all. It is this that leads to a nonlocal kernel that is not sensitive to the variability of the target \cite{Bengio06thecurse} (see Section \ref{sec:results:fz}).

$\bm{\theta}$ and $\sigma_n^2$ were adjusted to maximize the leave-one-out (LOO) log predictive probability \cite{Rasmussen:2005:GPM:1162254}:
\begin{equation}
    \label{eq:L_LOO}
    L_\text{LOO}(\mathcal{D}_\text{l},\mathbf{y},\bm{\theta},\sigma_n^2) = \sum_i^{M_\text{l}} \log p(y_i|\mathcal{D}_{\text{l}-i},\mathbf{y}_{\text{l}-i},\bm{\theta},\sigma_n^2)
\end{equation}
where $\mathcal{D}_\text{l} = \{\mathbf{z}_i\}_{i=1}^{M_\text{l}}$ is the set of (labeled) input data and:
\begin{equation*}
    \label{eq:log_p}
    \log p(y_i|\mathcal{D}_{\text{l}-i},\mathbf{y}_{\text{l}-i},\bm{\theta},\sigma_n^2) = -\frac{1}{2} \log \sigma_i^2 - \frac{(y_i - \mu_i)^2}{2 \sigma_i^2} - \frac{1}{2} \log 2 \pi
\end{equation*}
is the predictive log probability of the dataset $(\mathcal{D}_{\text{l}-i},\mathbf{y}_{\text{l}-i})$, formed by leaving out training case $i$, where:
\begin{equation*}
    \label{eq:log_p_mu}
    \mu_i = y_i - \frac{[(\mathbf{K} + \sigma_n^2 \mathbf{I})^{-1}\mathbf{y}]_i}{[(\mathbf{K} + \sigma_n^2 \mathbf{I})^{-1}]_{ii}}
\end{equation*}
\begin{equation*}
    \label{eq:log_p_sigma}
    \sigma_i^2 = \frac{1}{[(\mathbf{K} + \sigma_n^2 \mathbf{I})^{-1}]_{ii}}
\end{equation*}
are the predictive mean and variance. This was accomplished by minimizing the negative of Eq.\ (\ref{eq:L_LOO}), using simulated annealing \cite{Corana:1987:MMF:29380.29864}.


Following training, predictions from the DBN+GP model are made as follows: Given a test vector $\mathbf{v}_*$, its feature vector $\mathbf{z}_*$ is first calculated by the mapping $F$. Then, a GP prediction is obtained by conditioning on the (labeled) training data and $\bm{\theta}$. The distribution of the predicted value $y_*$ at $\mathbf{z}_*$ ($\mathbf{v}_*$) is:
\begin{equation}
    \label{eq:GP_py}
    p(y_*|\mathbf{z}_*,\mathcal{D}_\text{l},\bm{\theta},\sigma_n^2) = \mathcal{N}(y_*|\mathbf{k}^T_*(\mathbf{K} + \sigma_n^2 \mathbf{I})^{-1}\mathbf{y}, \mathbf{k}_{**} - \mathbf{k}_*^T(\mathbf{K} + \sigma_n^2 \mathbf{I})^{-1}\mathbf{k}_*)
\end{equation}
where $\mathbf{k}_* = \mathbf{k}(\mathbf{z}_*,\mathbf{Z}_\text{l})$ and $\mathbf{k}_{**} = k(\mathbf{z}_*,\mathbf{z}_*)$.

\section{Model system}
\label{sec:methods:model}

Analogous to Ref.\ \onlinecite{PhysRevLett.108.253002}, continuous potentials $v(x)$ for the model system described in Section \ref{sec:results:model} were randomly generated from:
\begin{equation}
    \label{eq:vx}
    v(x) = - \sum_{i=1}^3 a_i \exp[-(x - b_i)^2/(2 c_i^2)]
\end{equation}
where $a_i$, $b_i$, and $c_i$ were selected uniformly over $1 < a < 10$, $0.4 < b < 0.6$, and $0.03 < c < 0.1$. Hard walls were placed at $x = 0$ and $1$.

The Schr\"{o}dinger equation was solved numerically for $\{\phi_i\}_{i=1}^N$ and their corresponding energies $\{\epsilon_i\}_{i=1}^N$, by discretizing the domain using $n_x = 100$ grid points and using Numerov's method in matrix form \cite{:/content/aapt/journal/ajp/80/11/10.1119/1.4748813}. From these:
\begin{equation}
    \label{eq:En}
    E[n] = \sum_{i=1}^N \epsilon_i
\end{equation}
and using Eq.\ (\ref{eq:n}):
\begin{equation}
    \label{eq:n_numeric}
    n(x_i) = \sum_{i=1}^N | \phi_i(x_i) |^2 ~~~ , ~~~ n = \sum_{i=1}^{n_x} \Delta x ~ n(x_i)
\end{equation}
where $\Delta x = 1/2(n_x-1)$ if $i = 1$ or $n_x$, or $\Delta x = 1/(n_x-1)$ otherwise. From Eqs.\ (\ref{eq:HK})--(\ref{eq:KS2}):
\begin{equation}
    \label{eq:T_numeric}
    T_s[n] = E[n] - V[n]
\end{equation}
\begin{equation}
    \label{eq:V_numeric}
    V[n] = \sum_{i=1}^{n_x} \Delta x ~ n(x_i) v(x_i)  
\end{equation}
(both $E_\text{H}[n]$ and $E_\text{xc}[n]$ are zero for noninteracting electrons).

The above procedure was used to generate a data set consisting of $6000$ (${\bm{n}, T_s[n]}$) data points, where $\bm{n} = [n(x_1) ~ n(x_2) ~ \dots ~ n(x_{n_x})]$. In order to minimize possible bias in sampled data (which is especially important for small samples), data points were selected randomly from the data set without replacement (typically, $25\%$ in total); results were then obtained as averages over several samplings; discussed further in Appendix \ref{sec:methods:statistics}.

\section{Performance evaluation}
\label{sec:methods:statistics}

Equations for the NMSE, NMBF, and $r^2$ are:
%
\begin{equation}
    \label{eq:NMSE}
    \text{NMSE} = \overline{(y_* - y)^2}/(\overline{y_*} ~ \overline{y})
\end{equation}
%
%
\begin{equation}
    \label{eq:NMBF}
    \text{NMBF} = \begin{cases}
                             \overline{y_*}/\overline{y} - 1  & \overline{y_*} \ge \overline{y} \\
                             1 - \overline{y}/\overline{y_*} & \overline{y_*} < \overline{y}
    \end{cases}
\end{equation}
%
%
\begin{equation}
    \label{eq:r2}
    r^2 = \text{ss}^2_{y_* y}/(\text{ss}_{y_* y_*} \text{ss}_{y y})
\end{equation}
respectively, where $y = T_s[n]$ and $y_*$ is the corresponding DBN+GP prediction, and in Eq.\ (\ref{eq:r2}), $\text{ss}$ are the (unnormalized) covariance and variances of $y$ and $y_*$.

In the calculation of Eqs.\ (\ref{eq:NMSE})--(\ref{eq:r2}), there are two types of uncertainty. Consider training a single model. Testing it on unseen data provides the information necessary to estimate these quantities, as well as the model uncertainty; $1000$ data points were used. This model, however, is parameterized by $\mathbf{W}$, $\bm{\theta}$, and $\sigma_n^2$, which are determined by stochastic methods and using randomly-sampled training data. This leads to parameter uncertainty. This can be determined by training and testing several models; $10$ were used.

Model and parameter uncertainties are both informative, but useful for different purposes. The latter, for example, is necessary in order to make meaningful comparisons among model details; this is therefore the type shown in Tables \ref{Tb:architecture}--\ref{Tb:labled_data}. The former is shown in Tables \ref{Tb:n} and \ref{Tb:SCn}. Comparing these results shows that the magnitudes of the two types of uncertainty are similar.

In order to determine both types of uncertainty and also correct any bias in the estimation of Eqs.\ (\ref{eq:NMSE})--(\ref{eq:r2}), bootstrap resampling \cite{efron1979} was used; $100000$ samplings were made. 

\section{Self-consistent densities}
\label{sec:methods:SCn}
%
%

In Kohn--Sham density-functional theory \cite{RevModPhys.87.897}, minimization of the energy functional in Eq.\ (\ref{eq:HK}) is typically performed by a self-consistent procedure that requires calculating the variation of the energy functional with respect to the density. Calculating a stable estimate of a functional derivative using (certain) machine learning algorithms can be non-trivial though; this was shown in Ref.\ \onlinecite{kernel_functionalDerivative_problem} for kernel methods, for example. Self-consistent densities were therefore obtained by searching for a density which minimizes Eq.\ \eqref{eq:HK}, with the addition of a penalty term to conserve the number of electrons:
\begin{equation}
    \label{eq:fSCn}
    E[n] = y_* + V[n] + \frac{1}{2 \mu} \left( n - N \right)^2
\end{equation}
where $\mu$ is the penalty factor and other quantities have been defined previously. For the results in Table \ref{Tb:SCn}, trial and error suggested $\mu = 10^{-5}$ as a reasonable choice. This search was performed stochastically, using simulated annealing \cite{Corana:1987:MMF:29380.29864} (see also below). Note that the initial density was taken to be the mean as calculated from the training set; and during annealing, the density was constrained to lie within its bounds. 

Initial calculations suggest that the use of a DBN to initialize a supervised learning algorithm (i.e., the approach developed in this work) may be capable of calculating stable and accurate functional derivatives. The application domain thus far has been limited to toy mathematical problems, and so it is too early to tell whether this will work for density functionals. While a complete discussion of this is beyond the scope of this work, these general findings can be qualitatively understood as follows: If the underlying dimensionality of the data is less than that of the input domain, then conventional machine learning will be unable to describe a functional derivative; no data exists along the extraneous (and orthogonal) dimensions. Previous approaches have been based on minimization in a projected subspace; for example, linear principal component analysis in Ref.\ \onlinecite{PhysRevLett.108.253002} and a nonlinear approach in Ref.\ \onlinecite{QUA:QUA24937}. There is no guarantee though that such projected dimensions describe inherent features of the data. Moreover, it has been shown \cite{Hinton504} that a deep type of neural network works much better to (naturally) reduce the dimensionality of data, provided that its weights have been effectively initialized (e.g., by a DBN). This approach may improve the already accurate results in Table \ref{Tb:SCn}, with no increase in computational cost.   

\section{Generative sampling}
\label{sec:methods:gen_sampling}

Generative sampling can be used to draw samples from $p(\mathbf{v};\mathbf{W})$. This is accomplished by setting up a Markov chain that converges to $p(\mathbf{v};\mathbf{W})$, and running it to equilibrium. In practice, random states for the visible units of the top-level RBM are assigned, and iteratively sampling of $\mathbf{h}$ and $\mathbf{v}$ is performed. After this achieves equilibrium, the resulting visible units (of this RBM) are deterministically backpropagated.

For the results in Section \ref{sec:results:gen_sampling}, $100000$ steps of sampling were performed. A mean-field approximation (discussed in Appendix \ref{sec:methods:ML_alg}) was made for the visible units.


\bibliography{/Users/jmcmahon/Desktop/Manuscripts/Bibtex_Refs/journal_names_s,./manuscript}

\begin{thebibliography}{43}%
\makeatletter
\providecommand \@ifxundefined [1]{%
 \@ifx{#1\undefined}
}%
\providecommand \@ifnum [1]{%
 \ifnum #1\expandafter \@firstoftwo
 \else \expandafter \@secondoftwo
 \fi
}%
\providecommand \@ifx [1]{%
 \ifx #1\expandafter \@firstoftwo
 \else \expandafter \@secondoftwo
 \fi
}%
\providecommand \natexlab [1]{#1}%
\providecommand \enquote  [1]{``#1''}%
\providecommand \bibnamefont  [1]{#1}%
\providecommand \bibfnamefont [1]{#1}%
\providecommand \citenamefont [1]{#1}%
\providecommand \href@noop [0]{\@secondoftwo}%
\providecommand \href [0]{\begingroup \@sanitize@url \@href}%
\providecommand \@href[1]{\@@startlink{#1}\@@href}%
\providecommand \@@href[1]{\endgroup#1\@@endlink}%
\providecommand \@sanitize@url [0]{\catcode `\\12\catcode `\$12\catcode
  `\&12\catcode `\#12\catcode `\^12\catcode `\_12\catcode `\%12\relax}%
\providecommand \@@startlink[1]{}%
\providecommand \@@endlink[0]{}%
\providecommand \url  [0]{\begingroup\@sanitize@url \@url }%
\providecommand \@url [1]{\endgroup\@href {#1}{\urlprefix }}%
\providecommand \urlprefix  [0]{URL }%
\providecommand \Eprint [0]{\href }%
\providecommand \doibase [0]{http://dx.doi.org/}%
\providecommand \selectlanguage [0]{\@gobble}%
\providecommand \bibinfo  [0]{\@secondoftwo}%
\providecommand \bibfield  [0]{\@secondoftwo}%
\providecommand \translation [1]{[#1]}%
\providecommand \BibitemOpen [0]{}%
\providecommand \bibitemStop [0]{}%
\providecommand \bibitemNoStop [0]{.\EOS\space}%
\providecommand \EOS [0]{\spacefactor3000\relax}%
\providecommand \BibitemShut  [1]{\csname bibitem#1\endcsname}%
\let\auto@bib@innerbib\@empty
\bibitem [{\citenamefont {Hohenberg}\ and\ \citenamefont
  {Kohn}(1964)}]{PhysRev.136.B864}%
  \BibitemOpen
  \bibfield  {author} {\bibinfo {author} {\bibfnamefont {P.}~\bibnamefont
  {Hohenberg}}\ and\ \bibinfo {author} {\bibfnamefont {W.}~\bibnamefont
  {Kohn}},\ }\href {\doibase 10.1103/PhysRev.136.B864} {\bibfield  {journal}
  {\bibinfo  {journal} {Phys. Rev.}\ }\textbf {\bibinfo {volume} {136}},\
  \bibinfo {pages} {B864} (\bibinfo {year} {1964})}\BibitemShut {NoStop}%
\bibitem [{\citenamefont {Kohn}\ and\ \citenamefont
  {Sham}(1965)}]{PhysRev.140.A1133}%
  \BibitemOpen
  \bibfield  {author} {\bibinfo {author} {\bibfnamefont {W.}~\bibnamefont
  {Kohn}}\ and\ \bibinfo {author} {\bibfnamefont {L.~J.}\ \bibnamefont
  {Sham}},\ }\href {\doibase 10.1103/PhysRev.140.A1133} {\bibfield  {journal}
  {\bibinfo  {journal} {Phys. Rev.}\ }\textbf {\bibinfo {volume} {140}},\
  \bibinfo {pages} {A1133} (\bibinfo {year} {1965})}\BibitemShut {NoStop}%
\bibitem [{\citenamefont {Jones}(2015)}]{RevModPhys.87.897}%
  \BibitemOpen
  \bibfield  {author} {\bibinfo {author} {\bibfnamefont {R.~O.}\ \bibnamefont
  {Jones}},\ }\href {\doibase 10.1103/RevModPhys.87.897} {\bibfield  {journal}
  {\bibinfo  {journal} {Rev. Mod. Phys.}\ }\textbf {\bibinfo {volume} {87}},\
  \bibinfo {pages} {897} (\bibinfo {year} {2015})}\BibitemShut {NoStop}%
\bibitem [{\citenamefont
  {Burke}(2012)}]{:/content/aip/journal/jcp/136/15/10.1063/1.4704546}%
  \BibitemOpen
  \bibfield  {author} {\bibinfo {author} {\bibfnamefont {K.}~\bibnamefont
  {Burke}},\ }\href {\doibase http://dx.doi.org/10.1063/1.4704546} {\bibfield
  {journal} {\bibinfo  {journal} {The Journal of Chemical Physics}\ }\textbf
  {\bibinfo {volume} {136}},\ \bibinfo {eid} {150901} (\bibinfo {year}
  {2012}),\ http://dx.doi.org/10.1063/1.4704546}\BibitemShut {NoStop}%
\bibitem [{\citenamefont {Karasiev}\ and\ \citenamefont
  {Trickey}(2012)}]{Karasiev20122519}%
  \BibitemOpen
  \bibfield  {author} {\bibinfo {author} {\bibfnamefont {V.}~\bibnamefont
  {Karasiev}}\ and\ \bibinfo {author} {\bibfnamefont {S.}~\bibnamefont
  {Trickey}},\ }\href {\doibase http://dx.doi.org/10.1016/j.cpc.2012.06.016}
  {\bibfield  {journal} {\bibinfo  {journal} {Computer Physics Communications}\
  }\textbf {\bibinfo {volume} {183}},\ \bibinfo {pages} {2519 } (\bibinfo
  {year} {2012})}\BibitemShut {NoStop}%
\bibitem [{\citenamefont {Perdew}\ \emph {et~al.}(1996)\citenamefont {Perdew},
  \citenamefont {Burke},\ and\ \citenamefont
  {Ernzerhof}}]{PhysRevLett.77.3865}%
  \BibitemOpen
  \bibfield  {author} {\bibinfo {author} {\bibfnamefont {J.~P.}\ \bibnamefont
  {Perdew}}, \bibinfo {author} {\bibfnamefont {K.}~\bibnamefont {Burke}}, \
  and\ \bibinfo {author} {\bibfnamefont {M.}~\bibnamefont {Ernzerhof}},\ }\href
  {\doibase 10.1103/PhysRevLett.77.3865} {\bibfield  {journal} {\bibinfo
  {journal} {Phys. Rev. Lett.}\ }\textbf {\bibinfo {volume} {77}},\ \bibinfo
  {pages} {3865} (\bibinfo {year} {1996})}\BibitemShut {NoStop}%
\bibitem [{\citenamefont {Becke}(1988)}]{PhysRevA.38.3098}%
  \BibitemOpen
  \bibfield  {author} {\bibinfo {author} {\bibfnamefont {A.~D.}\ \bibnamefont
  {Becke}},\ }\href {\doibase 10.1103/PhysRevA.38.3098} {\bibfield  {journal}
  {\bibinfo  {journal} {Phys. Rev. A}\ }\textbf {\bibinfo {volume} {38}},\
  \bibinfo {pages} {3098} (\bibinfo {year} {1988})}\BibitemShut {NoStop}%
\bibitem [{\citenamefont {Lee}\ \emph {et~al.}(1988)\citenamefont {Lee},
  \citenamefont {Yang},\ and\ \citenamefont {Parr}}]{PhysRevB.37.785}%
  \BibitemOpen
  \bibfield  {author} {\bibinfo {author} {\bibfnamefont {C.}~\bibnamefont
  {Lee}}, \bibinfo {author} {\bibfnamefont {W.}~\bibnamefont {Yang}}, \ and\
  \bibinfo {author} {\bibfnamefont {R.~G.}\ \bibnamefont {Parr}},\ }\href
  {\doibase 10.1103/PhysRevB.37.785} {\bibfield  {journal} {\bibinfo  {journal}
  {Phys. Rev. B}\ }\textbf {\bibinfo {volume} {37}},\ \bibinfo {pages} {785}
  (\bibinfo {year} {1988})}\BibitemShut {NoStop}%
\bibitem [{\citenamefont {Snyder}\ \emph {et~al.}(2012)\citenamefont {Snyder},
  \citenamefont {Rupp}, \citenamefont {Hansen}, \citenamefont {M\"uller},\ and\
  \citenamefont {Burke}}]{PhysRevLett.108.253002}%
  \BibitemOpen
  \bibfield  {author} {\bibinfo {author} {\bibfnamefont {J.~C.}\ \bibnamefont
  {Snyder}}, \bibinfo {author} {\bibfnamefont {M.}~\bibnamefont {Rupp}},
  \bibinfo {author} {\bibfnamefont {K.}~\bibnamefont {Hansen}}, \bibinfo
  {author} {\bibfnamefont {K.-R.}\ \bibnamefont {M\"uller}}, \ and\ \bibinfo
  {author} {\bibfnamefont {K.}~\bibnamefont {Burke}},\ }\href {\doibase
  10.1103/PhysRevLett.108.253002} {\bibfield  {journal} {\bibinfo  {journal}
  {Phys. Rev. Lett.}\ }\textbf {\bibinfo {volume} {108}},\ \bibinfo {pages}
  {253002} (\bibinfo {year} {2012})}\BibitemShut {NoStop}%
\bibitem [{\citenamefont {Duda}\ and\ \citenamefont {Hart}(1973)}]{Duda_Hart}%
  \BibitemOpen
  \bibfield  {author} {\bibinfo {author} {\bibfnamefont {R.~O.}\ \bibnamefont
  {Duda}}\ and\ \bibinfo {author} {\bibfnamefont {P.~E.}\ \bibnamefont
  {Hart}},\ }\href@noop {} {\emph {\bibinfo {title} {Pattern Classification and
  Scene Analysis}}}\ (\bibinfo  {publisher} {Wiley},\ \bibinfo {year}
  {1973})\BibitemShut {NoStop}%
\bibitem [{\citenamefont {Snyder}\ \emph
  {et~al.}(2013{\natexlab{a}})\citenamefont {Snyder}, \citenamefont {Rupp},
  \citenamefont {Hansen}, \citenamefont {Blooston}, \citenamefont {MŸller},\
  and\ \citenamefont
  {Burke}}]{:/content/aip/journal/jcp/139/22/10.1063/1.4834075}%
  \BibitemOpen
  \bibfield  {author} {\bibinfo {author} {\bibfnamefont {J.~C.}\ \bibnamefont
  {Snyder}}, \bibinfo {author} {\bibfnamefont {M.}~\bibnamefont {Rupp}},
  \bibinfo {author} {\bibfnamefont {K.}~\bibnamefont {Hansen}}, \bibinfo
  {author} {\bibfnamefont {L.}~\bibnamefont {Blooston}}, \bibinfo {author}
  {\bibfnamefont {K.-R.}\ \bibnamefont {MŸller}}, \ and\ \bibinfo {author}
  {\bibfnamefont {K.}~\bibnamefont {Burke}},\ }\href {\doibase
  http://dx.doi.org/10.1063/1.4834075} {\bibfield  {journal} {\bibinfo
  {journal} {The Journal of Chemical Physics}\ }\textbf {\bibinfo {volume}
  {139}},\ \bibinfo {eid} {224104} (\bibinfo {year}
  {2013}{\natexlab{a}})}\BibitemShut {NoStop}%
\bibitem [{\citenamefont {Sch\"{o}lkopf}\ and\ \citenamefont
  {Smola}(2001)}]{Learning_with_Kernels}%
  \BibitemOpen
  \bibfield  {author} {\bibinfo {author} {\bibfnamefont {B.}~\bibnamefont
  {Sch\"{o}lkopf}}\ and\ \bibinfo {author} {\bibfnamefont {A.~J.}\ \bibnamefont
  {Smola}},\ }\href@noop {} {\emph {\bibinfo {title} {Learning with Kernels}}}\
  (\bibinfo  {publisher} {The MIT Press},\ \bibinfo {year} {2001})\BibitemShut
  {NoStop}%
\bibitem [{\citenamefont {Li}\ \emph {et~al.}(2016)\citenamefont {Li},
  \citenamefont {Snyder}, \citenamefont {Pelaschier}, \citenamefont {Huang},
  \citenamefont {Niranjan}, \citenamefont {Duncan}, \citenamefont {Rupp},
  \citenamefont {MŸller},\ and\ \citenamefont {Burke}}]{QUA:QUA25040}%
  \BibitemOpen
  \bibfield  {author} {\bibinfo {author} {\bibfnamefont {L.}~\bibnamefont
  {Li}}, \bibinfo {author} {\bibfnamefont {J.~C.}\ \bibnamefont {Snyder}},
  \bibinfo {author} {\bibfnamefont {I.~M.}\ \bibnamefont {Pelaschier}},
  \bibinfo {author} {\bibfnamefont {J.}~\bibnamefont {Huang}}, \bibinfo
  {author} {\bibfnamefont {U.-N.}\ \bibnamefont {Niranjan}}, \bibinfo {author}
  {\bibfnamefont {P.}~\bibnamefont {Duncan}}, \bibinfo {author} {\bibfnamefont
  {M.}~\bibnamefont {Rupp}}, \bibinfo {author} {\bibfnamefont {K.-R.}\
  \bibnamefont {MŸller}}, \ and\ \bibinfo {author} {\bibfnamefont
  {K.}~\bibnamefont {Burke}},\ }\href {\doibase 10.1002/qua.25040} {\bibfield
  {journal} {\bibinfo  {journal} {International Journal of Quantum Chemistry}\
  }\textbf {\bibinfo {volume} {116}},\ \bibinfo {pages} {819} (\bibinfo {year}
  {2016})}\BibitemShut {NoStop}%
\bibitem [{\citenamefont {Bengio}\ \emph {et~al.}(2006)\citenamefont {Bengio},
  \citenamefont {Delalleau},\ and\ \citenamefont {Roux}}]{Bengio06thecurse}%
  \BibitemOpen
  \bibfield  {author} {\bibinfo {author} {\bibfnamefont {Y.}~\bibnamefont
  {Bengio}}, \bibinfo {author} {\bibfnamefont {O.}~\bibnamefont {Delalleau}}, \
  and\ \bibinfo {author} {\bibfnamefont {N.~L.}\ \bibnamefont {Roux}},\ }in\
  \href@noop {} {\emph {\bibinfo {booktitle} {In Advances in Neural Information
  Processing Systems 18}}}\ (\bibinfo  {publisher} {MIT Press},\ \bibinfo
  {year} {2006})\ p.\ \bibinfo {pages} {2006}\BibitemShut {NoStop}%
\bibitem [{\citenamefont {LeCun}\ \emph {et~al.}(2015)\citenamefont {LeCun},
  \citenamefont {Bengio},\ and\ \citenamefont {Hinton}}]{Nature.521.436}%
  \BibitemOpen
  \bibfield  {author} {\bibinfo {author} {\bibfnamefont {Y.}~\bibnamefont
  {LeCun}}, \bibinfo {author} {\bibfnamefont {Y.}~\bibnamefont {Bengio}}, \
  and\ \bibinfo {author} {\bibfnamefont {G.}~\bibnamefont {Hinton}},\ }\href
  {\doibase 10.1038/nature14539} {\bibfield  {journal} {\bibinfo  {journal}
  {Nature}\ }\textbf {\bibinfo {volume} {521}},\ \bibinfo {pages} {436}
  (\bibinfo {year} {2015})}\BibitemShut {NoStop}%
\bibitem [{\citenamefont {Erhan}\ \emph {et~al.}(2010)\citenamefont {Erhan},
  \citenamefont {Courville},\ and\ \citenamefont
  {Bengio}}]{Erhan-vis-techreport-2010}%
  \BibitemOpen
  \bibfield  {author} {\bibinfo {author} {\bibfnamefont {D.}~\bibnamefont
  {Erhan}}, \bibinfo {author} {\bibfnamefont {A.}~\bibnamefont {Courville}}, \
  and\ \bibinfo {author} {\bibfnamefont {Y.}~\bibnamefont {Bengio}},\
  }\href@noop {} {\emph {\bibinfo {title} {Understanding Representations
  Learned in Deep Architectures}}},\ \bibinfo {type} {Tech. Rep.}\ \bibinfo
  {number} {1355}\ (\bibinfo  {institution} {Universit{\'{e}} de
  Montr{\'{e}}al/DIRO},\ \bibinfo {year} {2010})\BibitemShut {NoStop}%
\bibitem [{\citenamefont {Goodfellow}\ \emph {et~al.}(2009)\citenamefont
  {Goodfellow}, \citenamefont {Lee}, \citenamefont {Le}, \citenamefont {Saxe},\
  and\ \citenamefont {Ng}}]{NIPS2009_3790}%
  \BibitemOpen
  \bibfield  {author} {\bibinfo {author} {\bibfnamefont {I.}~\bibnamefont
  {Goodfellow}}, \bibinfo {author} {\bibfnamefont {H.}~\bibnamefont {Lee}},
  \bibinfo {author} {\bibfnamefont {Q.~V.}\ \bibnamefont {Le}}, \bibinfo
  {author} {\bibfnamefont {A.}~\bibnamefont {Saxe}}, \ and\ \bibinfo {author}
  {\bibfnamefont {A.~Y.}\ \bibnamefont {Ng}},\ }in\ \href
  {http://papers.nips.cc/paper/3790-measuring-invariances-in-deep-networks.pdf}
  {\emph {\bibinfo {booktitle} {Advances in Neural Information Processing
  Systems 22}}},\ \bibinfo {editor} {edited by\ \bibinfo {editor}
  {\bibfnamefont {Y.}~\bibnamefont {Bengio}}, \bibinfo {editor} {\bibfnamefont
  {D.}~\bibnamefont {Schuurmans}}, \bibinfo {editor} {\bibfnamefont
  {J.}~\bibnamefont {Lafferty}}, \bibinfo {editor} {\bibfnamefont
  {C.}~\bibnamefont {Williams}}, \ and\ \bibinfo {editor} {\bibfnamefont
  {A.}~\bibnamefont {Culotta}}}\ (\bibinfo  {publisher} {Curran Associates,
  Inc.},\ \bibinfo {year} {2009})\ pp.\ \bibinfo {pages} {646--654}\BibitemShut
  {NoStop}%
\bibitem [{\citenamefont {Deng}(2014)}]{SIP:9155271}%
  \BibitemOpen
  \bibfield  {author} {\bibinfo {author} {\bibfnamefont {L.}~\bibnamefont
  {Deng}},\ }\href {\doibase 10.1017/atsip.2013.9} {\bibfield  {journal}
  {\bibinfo  {journal} {APSIPA Transactions on Signal and Information
  Processing}\ }\textbf {\bibinfo {volume} {3}} (\bibinfo {year} {2014}),\
  10.1017/atsip.2013.9}\BibitemShut {NoStop}%
\bibitem [{Note1()}]{Note1}%
  \BibitemOpen
  \bibinfo {note} {In this work, the input data $\protect \mathbf {v}$ is
  continuous. Therefore, $p(\protect \mathbf {v})$ corresponds to a probability
  density function. The notation was chosen to be consistent with that commonly
  used.}\BibitemShut {Stop}%
\bibitem [{\citenamefont {McQuarrie}(2000)}]{mcquarrie2000statistical}%
  \BibitemOpen
  \bibfield  {author} {\bibinfo {author} {\bibfnamefont {D.}~\bibnamefont
  {McQuarrie}},\ }\href {https://books.google.com/books?id=itcpPnDnJM0C} {\emph
  {\bibinfo {title} {Statistical Mechanics}}}\ (\bibinfo  {publisher}
  {University Science Books},\ \bibinfo {year} {2000})\BibitemShut {NoStop}%
\bibitem [{\citenamefont {Hopfield}(1982)}]{Hopfield01041982}%
  \BibitemOpen
  \bibfield  {author} {\bibinfo {author} {\bibfnamefont {J.~J.}\ \bibnamefont
  {Hopfield}},\ }\href {http://www.pnas.org/content/79/8/2554.abstract}
  {\bibfield  {journal} {\bibinfo  {journal} {Proceedings of the National
  Academy of Sciences}\ }\textbf {\bibinfo {volume} {79}},\ \bibinfo {pages}
  {2554} (\bibinfo {year} {1982})},\ \Eprint
  {http://arxiv.org/abs/http://www.pnas.org/content/79/8/2554.full.pdf}
  {http://www.pnas.org/content/79/8/2554.full.pdf} \BibitemShut {NoStop}%
\bibitem [{\citenamefont {Hanna}\ and\ \citenamefont
  {Heinold}(1985)}]{Hanna_NMSE_1985}%
  \BibitemOpen
  \bibfield  {author} {\bibinfo {author} {\bibfnamefont {S.~R.}\ \bibnamefont
  {Hanna}}\ and\ \bibinfo {author} {\bibfnamefont {D.~W.}\ \bibnamefont
  {Heinold}},\ }\href@noop {} {\emph {\bibinfo {title} {Development and
  Application of a Simple Method for Evaluating Air Quality Models}}},\
  \bibinfo {type} {Tech. Rep.}\ \bibinfo {number} {API Publication No.\ 4409}\
  (\bibinfo  {institution} {American Petroleum Institute},\ \bibinfo {address}
  {Washington, DC},\ \bibinfo {year} {1985})\BibitemShut {NoStop}%
\bibitem [{\citenamefont {Yu}\ \emph {et~al.}(2006)\citenamefont {Yu},
  \citenamefont {Eder}, \citenamefont {Dennis}, \citenamefont {Chu},\ and\
  \citenamefont {Schwartz}}]{ASL:ASL125}%
  \BibitemOpen
  \bibfield  {author} {\bibinfo {author} {\bibfnamefont {S.}~\bibnamefont
  {Yu}}, \bibinfo {author} {\bibfnamefont {B.}~\bibnamefont {Eder}}, \bibinfo
  {author} {\bibfnamefont {R.}~\bibnamefont {Dennis}}, \bibinfo {author}
  {\bibfnamefont {S.-H.}\ \bibnamefont {Chu}}, \ and\ \bibinfo {author}
  {\bibfnamefont {S.~E.}\ \bibnamefont {Schwartz}},\ }\href {\doibase
  10.1002/asl.125} {\bibfield  {journal} {\bibinfo  {journal} {Atmospheric
  Science Letters}\ }\textbf {\bibinfo {volume} {7}},\ \bibinfo {pages} {26}
  (\bibinfo {year} {2006})}\BibitemShut {NoStop}%
\bibitem [{\citenamefont {Pearson}(1895)}]{Pearson01011895}%
  \BibitemOpen
  \bibfield  {author} {\bibinfo {author} {\bibfnamefont {K.}~\bibnamefont
  {Pearson}},\ }\href {\doibase 10.1098/rspl.1895.0041} {\bibfield  {journal}
  {\bibinfo  {journal} {Proceedings of the Royal Society of London}\ }\textbf
  {\bibinfo {volume} {58}},\ \bibinfo {pages} {240} (\bibinfo {year} {1895})},\
  \Eprint
  {http://arxiv.org/abs/http://rspl.royalsocietypublishing.org/content/58/347-352/240.full.pdf+html}
  {http://rspl.royalsocietypublishing.org/content/58/347-352/240.full.pdf+html}
  \BibitemShut {NoStop}%
\bibitem [{\citenamefont {Elliott}\ \emph {et~al.}(2008)\citenamefont
  {Elliott}, \citenamefont {Lee}, \citenamefont {Cangi},\ and\ \citenamefont
  {Burke}}]{PhysRevLett.100.256406}%
  \BibitemOpen
  \bibfield  {author} {\bibinfo {author} {\bibfnamefont {P.}~\bibnamefont
  {Elliott}}, \bibinfo {author} {\bibfnamefont {D.}~\bibnamefont {Lee}},
  \bibinfo {author} {\bibfnamefont {A.}~\bibnamefont {Cangi}}, \ and\ \bibinfo
  {author} {\bibfnamefont {K.}~\bibnamefont {Burke}},\ }\href {\doibase
  10.1103/PhysRevLett.100.256406} {\bibfield  {journal} {\bibinfo  {journal}
  {Phys. Rev. Lett.}\ }\textbf {\bibinfo {volume} {100}},\ \bibinfo {pages}
  {256406} (\bibinfo {year} {2008})}\BibitemShut {NoStop}%
\bibitem [{\citenamefont {Le~Roux}\ and\ \citenamefont
  {Bengio}(2008)}]{LeRoux:2008:RPR:1374176.1374187}%
  \BibitemOpen
  \bibfield  {author} {\bibinfo {author} {\bibfnamefont {N.}~\bibnamefont
  {Le~Roux}}\ and\ \bibinfo {author} {\bibfnamefont {Y.}~\bibnamefont
  {Bengio}},\ }\href {\doibase 10.1162/neco.2008.04-07-510} {\bibfield
  {journal} {\bibinfo  {journal} {Neural Comput.}\ }\textbf {\bibinfo {volume}
  {20}},\ \bibinfo {pages} {1631} (\bibinfo {year} {2008})}\BibitemShut
  {NoStop}%
\bibitem [{\citenamefont {Krause}\ \emph {et~al.}(2013)\citenamefont {Krause},
  \citenamefont {Fischer}, \citenamefont {Glasmachers},\ and\ \citenamefont
  {Igel}}]{ICML2013_krause13}%
  \BibitemOpen
  \bibfield  {author} {\bibinfo {author} {\bibfnamefont {O.}~\bibnamefont
  {Krause}}, \bibinfo {author} {\bibfnamefont {A.}~\bibnamefont {Fischer}},
  \bibinfo {author} {\bibfnamefont {T.}~\bibnamefont {Glasmachers}}, \ and\
  \bibinfo {author} {\bibfnamefont {C.}~\bibnamefont {Igel}},\ }in\ \href
  {http://jmlr.csail.mit.edu/proceedings/papers/v28/krause13.pdf} {\emph
  {\bibinfo {booktitle} {Proceedings of the 30th International Conference on
  Machine Learning (ICML-13)}}},\ Vol.\ \bibinfo {volume} {28(1)},\ \bibinfo
  {editor} {edited by\ \bibinfo {editor} {\bibfnamefont {S.}~\bibnamefont
  {Dasgupta}}\ and\ \bibinfo {editor} {\bibfnamefont {D.}~\bibnamefont
  {Mcallester}}}\ (\bibinfo  {publisher} {JMLR Workshop and Conference
  Proceedings},\ \bibinfo {year} {2013})\ pp.\ \bibinfo {pages}
  {419--426}\BibitemShut {NoStop}%
\bibitem [{\citenamefont {Bengio}\ \emph {et~al.}(2011)\citenamefont {Bengio},
  \citenamefont {Bastien}, \citenamefont {Bergeron}, \citenamefont
  {Boulanger-lewandowski}, \citenamefont {Breuel}, \citenamefont {Chherawala},
  \citenamefont {Cisse}, \citenamefont {C™tŽ}, \citenamefont {Erhan},
  \citenamefont {Eustache}, \citenamefont {Glorot}, \citenamefont {Muller},
  \citenamefont {Lebeuf}, \citenamefont {Pascanu}, \citenamefont {Rifai},
  \citenamefont {Savard},\ and\ \citenamefont
  {Sicard}}]{AISTATS2011_BengioBBBBCCCEEGMLPRSS11}%
  \BibitemOpen
  \bibfield  {author} {\bibinfo {author} {\bibfnamefont {Y.}~\bibnamefont
  {Bengio}}, \bibinfo {author} {\bibfnamefont {F.}~\bibnamefont {Bastien}},
  \bibinfo {author} {\bibfnamefont {A.}~\bibnamefont {Bergeron}}, \bibinfo
  {author} {\bibfnamefont {N.}~\bibnamefont {Boulanger-lewandowski}}, \bibinfo
  {author} {\bibfnamefont {T.~M.}\ \bibnamefont {Breuel}}, \bibinfo {author}
  {\bibfnamefont {Y.}~\bibnamefont {Chherawala}}, \bibinfo {author}
  {\bibfnamefont {M.}~\bibnamefont {Cisse}}, \bibinfo {author} {\bibfnamefont
  {M.}~\bibnamefont {C™tŽ}}, \bibinfo {author} {\bibfnamefont {D.}~\bibnamefont
  {Erhan}}, \bibinfo {author} {\bibfnamefont {J.}~\bibnamefont {Eustache}},
  \bibinfo {author} {\bibfnamefont {X.}~\bibnamefont {Glorot}}, \bibinfo
  {author} {\bibfnamefont {X.}~\bibnamefont {Muller}}, \bibinfo {author}
  {\bibfnamefont {S.~P.}\ \bibnamefont {Lebeuf}}, \bibinfo {author}
  {\bibfnamefont {R.}~\bibnamefont {Pascanu}}, \bibinfo {author} {\bibfnamefont
  {S.}~\bibnamefont {Rifai}}, \bibinfo {author} {\bibfnamefont
  {F.}~\bibnamefont {Savard}}, \ and\ \bibinfo {author} {\bibfnamefont
  {G.}~\bibnamefont {Sicard}},\ }in\ \href
  {http://www.jmlr.org/proceedings/papers/v15/bengio11b/bengio11b.pdf} {\emph
  {\bibinfo {booktitle} {Proceedings of the Fourteenth International Conference
  on Artificial Intelligence and Statistics (AISTATS-11)}}},\ Vol.~\bibinfo
  {volume} {15},\ \bibinfo {editor} {edited by\ \bibinfo {editor}
  {\bibfnamefont {G.~J.}\ \bibnamefont {Gordon}}\ and\ \bibinfo {editor}
  {\bibfnamefont {D.~B.}\ \bibnamefont {Dunson}}}\ (\bibinfo  {publisher}
  {Journal of Machine Learning Research - Workshop and Conference
  Proceedings},\ \bibinfo {year} {2011})\ pp.\ \bibinfo {pages}
  {164--172}\BibitemShut {NoStop}%
\bibitem [{\citenamefont {Ceperley}\ and\ \citenamefont
  {Alder}(1980)}]{PhysRevLett.45.566}%
  \BibitemOpen
  \bibfield  {author} {\bibinfo {author} {\bibfnamefont {D.~M.}\ \bibnamefont
  {Ceperley}}\ and\ \bibinfo {author} {\bibfnamefont {B.~J.}\ \bibnamefont
  {Alder}},\ }\href {\doibase 10.1103/PhysRevLett.45.566} {\bibfield  {journal}
  {\bibinfo  {journal} {Phys. Rev. Lett.}\ }\textbf {\bibinfo {volume} {45}},\
  \bibinfo {pages} {566} (\bibinfo {year} {1980})}\BibitemShut {NoStop}%
\bibitem [{\citenamefont {Nekovee}\ \emph {et~al.}(2001)\citenamefont
  {Nekovee}, \citenamefont {Foulkes},\ and\ \citenamefont
  {Needs}}]{PhysRevLett.87.036401}%
  \BibitemOpen
  \bibfield  {author} {\bibinfo {author} {\bibfnamefont {M.}~\bibnamefont
  {Nekovee}}, \bibinfo {author} {\bibfnamefont {W.~M.~C.}\ \bibnamefont
  {Foulkes}}, \ and\ \bibinfo {author} {\bibfnamefont {R.~J.}\ \bibnamefont
  {Needs}},\ }\href {\doibase 10.1103/PhysRevLett.87.036401} {\bibfield
  {journal} {\bibinfo  {journal} {Phys. Rev. Lett.}\ }\textbf {\bibinfo
  {volume} {87}},\ \bibinfo {pages} {036401} (\bibinfo {year}
  {2001})}\BibitemShut {NoStop}%
\bibitem [{\citenamefont {Muller}\ \emph {et~al.}(2016)\citenamefont {Muller},
  \citenamefont {Kusne},\ and\ \citenamefont {Ramprasad}}]{RevCompChem.29.4}%
  \BibitemOpen
  \bibfield  {author} {\bibinfo {author} {\bibfnamefont {T.}~\bibnamefont
  {Muller}}, \bibinfo {author} {\bibfnamefont {G.}~\bibnamefont {Kusne}}, \
  and\ \bibinfo {author} {\bibfnamefont {R.}~\bibnamefont {Ramprasad}},\ }in\
  \href@noop {} {\emph {\bibinfo {booktitle} {Rev. Comp. Chem. (accepted for
  publication)}}},\ Vol.~\bibinfo {volume} {29},\ \bibinfo {editor} {edited by\
  \bibinfo {editor} {\bibfnamefont {A.~L.}\ \bibnamefont {Parrill}}\ and\
  \bibinfo {editor} {\bibfnamefont {K.~B.}\ \bibnamefont {Lipkowitz}}}\
  (\bibinfo  {publisher} {John Wiley \& Sons, Inc.},\ \bibinfo {year}
  {2016})\BibitemShut {NoStop}%
\bibitem [{\citenamefont {Rupp}(2015)}]{QUA:QUA24954}%
  \BibitemOpen
  \bibfield  {author} {\bibinfo {author} {\bibfnamefont {M.}~\bibnamefont
  {Rupp}},\ }\href {\doibase 10.1002/qua.24954} {\bibfield  {journal} {\bibinfo
   {journal} {International Journal of Quantum Chemistry}\ }\textbf {\bibinfo
  {volume} {115}},\ \bibinfo {pages} {1058} (\bibinfo {year}
  {2015})}\BibitemShut {NoStop}%
\bibitem [{\citenamefont {Masters}(2015)}]{masters2015deep}%
  \BibitemOpen
  \bibfield  {author} {\bibinfo {author} {\bibfnamefont {T.}~\bibnamefont
  {Masters}},\ }\href@noop {} {\emph {\bibinfo {title} {Deep Belief Nets in C++
  and CUDA C: Volume I: Restricted Boltzmann Machines and Supervised
  Feedforward Networks}}},\ Deep Belief Nets in C++ and CUDA C\ (\bibinfo
  {publisher} {CreateSpace Independent Publishing Platform},\ \bibinfo {year}
  {2015})\BibitemShut {NoStop}%
\bibitem [{\citenamefont {Hinton}(2002)}]{Hinton:2002:TPE:639729.639730}%
  \BibitemOpen
  \bibfield  {author} {\bibinfo {author} {\bibfnamefont {G.~E.}\ \bibnamefont
  {Hinton}},\ }\href {\doibase 10.1162/089976602760128018} {\bibfield
  {journal} {\bibinfo  {journal} {Neural Comput.}\ }\textbf {\bibinfo {volume}
  {14}},\ \bibinfo {pages} {1771} (\bibinfo {year} {2002})}\BibitemShut
  {NoStop}%
\bibitem [{\citenamefont {Hinton}\ \emph {et~al.}(2006)\citenamefont {Hinton},
  \citenamefont {Osindero},\ and\ \citenamefont
  {Teh}}]{Hinton:2006:FLA:1161603.1161605}%
  \BibitemOpen
  \bibfield  {author} {\bibinfo {author} {\bibfnamefont {G.~E.}\ \bibnamefont
  {Hinton}}, \bibinfo {author} {\bibfnamefont {S.}~\bibnamefont {Osindero}}, \
  and\ \bibinfo {author} {\bibfnamefont {Y.-W.}\ \bibnamefont {Teh}},\ }\href
  {\doibase 10.1162/neco.2006.18.7.1527} {\bibfield  {journal} {\bibinfo
  {journal} {Neural Comput.}\ }\textbf {\bibinfo {volume} {18}},\ \bibinfo
  {pages} {1527} (\bibinfo {year} {2006})}\BibitemShut {NoStop}%
\bibitem [{\citenamefont {Hinton}\ and\ \citenamefont
  {Salakhutdinov}(2008)}]{NIPS2007_3211}%
  \BibitemOpen
  \bibfield  {author} {\bibinfo {author} {\bibfnamefont {G.~E.}\ \bibnamefont
  {Hinton}}\ and\ \bibinfo {author} {\bibfnamefont {R.~R.}\ \bibnamefont
  {Salakhutdinov}},\ }in\ \href
  {http://papers.nips.cc/paper/3211-using-deep-belief-nets-to-learn-covariance-kernels-for-gaussian-processes.pdf}
  {\emph {\bibinfo {booktitle} {Advances in Neural Information Processing
  Systems 20}}},\ \bibinfo {editor} {edited by\ \bibinfo {editor}
  {\bibfnamefont {J.}~\bibnamefont {Platt}}, \bibinfo {editor} {\bibfnamefont
  {D.}~\bibnamefont {Koller}}, \bibinfo {editor} {\bibfnamefont
  {Y.}~\bibnamefont {Singer}}, \ and\ \bibinfo {editor} {\bibfnamefont
  {S.}~\bibnamefont {Roweis}}}\ (\bibinfo  {publisher} {Curran Associates,
  Inc.},\ \bibinfo {year} {2008})\ pp.\ \bibinfo {pages}
  {1249--1256}\BibitemShut {NoStop}%
\bibitem [{\citenamefont {Rasmussen}\ and\ \citenamefont
  {Williams}(2005)}]{Rasmussen:2005:GPM:1162254}%
  \BibitemOpen
  \bibfield  {author} {\bibinfo {author} {\bibfnamefont {C.~E.}\ \bibnamefont
  {Rasmussen}}\ and\ \bibinfo {author} {\bibfnamefont {C.~K.~I.}\ \bibnamefont
  {Williams}},\ }\href@noop {} {\emph {\bibinfo {title} {Gaussian Processes for
  Machine Learning (Adaptive Computation and Machine Learning)}}}\ (\bibinfo
  {publisher} {The MIT Press},\ \bibinfo {year} {2005})\BibitemShut {NoStop}%
\bibitem [{\citenamefont {Corana}\ \emph {et~al.}(1987)\citenamefont {Corana},
  \citenamefont {Marchesi}, \citenamefont {Martini},\ and\ \citenamefont
  {Ridella}}]{Corana:1987:MMF:29380.29864}%
  \BibitemOpen
  \bibfield  {author} {\bibinfo {author} {\bibfnamefont {A.}~\bibnamefont
  {Corana}}, \bibinfo {author} {\bibfnamefont {M.}~\bibnamefont {Marchesi}},
  \bibinfo {author} {\bibfnamefont {C.}~\bibnamefont {Martini}}, \ and\
  \bibinfo {author} {\bibfnamefont {S.}~\bibnamefont {Ridella}},\ }\href
  {\doibase 10.1145/29380.29864} {\bibfield  {journal} {\bibinfo  {journal}
  {ACM Trans. Math. Softw.}\ }\textbf {\bibinfo {volume} {13}},\ \bibinfo
  {pages} {262} (\bibinfo {year} {1987})}\BibitemShut {NoStop}%
\bibitem [{\citenamefont {Pillai}\ \emph {et~al.}(2012)\citenamefont {Pillai},
  \citenamefont {Goglio},\ and\ \citenamefont
  {Walker}}]{:/content/aapt/journal/ajp/80/11/10.1119/1.4748813}%
  \BibitemOpen
  \bibfield  {author} {\bibinfo {author} {\bibfnamefont {M.}~\bibnamefont
  {Pillai}}, \bibinfo {author} {\bibfnamefont {J.}~\bibnamefont {Goglio}}, \
  and\ \bibinfo {author} {\bibfnamefont {T.~G.}\ \bibnamefont {Walker}},\
  }\href {\doibase http://dx.doi.org/10.1119/1.4748813} {\bibfield  {journal}
  {\bibinfo  {journal} {American Journal of Physics}\ }\textbf {\bibinfo
  {volume} {80}},\ \bibinfo {pages} {1017} (\bibinfo {year}
  {2012})}\BibitemShut {NoStop}%
\bibitem [{\citenamefont {Efron}(1979)}]{efron1979}%
  \BibitemOpen
  \bibfield  {author} {\bibinfo {author} {\bibfnamefont {B.}~\bibnamefont
  {Efron}},\ }\href {\doibase 10.1214/aos/1176344552} {\bibfield  {journal}
  {\bibinfo  {journal} {Ann. Statist.}\ }\textbf {\bibinfo {volume} {7}},\
  \bibinfo {pages} {1} (\bibinfo {year} {1979})}\BibitemShut {NoStop}%
\bibitem [{\citenamefont {Snyder}\ \emph
  {et~al.}(2013{\natexlab{b}})\citenamefont {Snyder}, \citenamefont
  {Sebastian}, \citenamefont {Burke},\ and\ \citenamefont
  {M{\"u}ller}}]{kernel_functionalDerivative_problem}%
  \BibitemOpen
  \bibfield  {author} {\bibinfo {author} {\bibfnamefont {J.~C.}\ \bibnamefont
  {Snyder}}, \bibinfo {author} {\bibfnamefont {M.}~\bibnamefont {Sebastian}},
  \bibinfo {author} {\bibfnamefont {K.}~\bibnamefont {Burke}}, \ and\ \bibinfo
  {author} {\bibfnamefont {K.-R.}\ \bibnamefont {M{\"u}ller}},\ }in\ \href@noop
  {} {\emph {\bibinfo {booktitle} {Empirical Inference}}},\ \bibinfo {editor}
  {edited by\ \bibinfo {editor} {\bibfnamefont {B.}~\bibnamefont
  {Sch{\"o}lkopf}}, \bibinfo {editor} {\bibfnamefont {Z.}~\bibnamefont {Luo}},
  \ and\ \bibinfo {editor} {\bibfnamefont {V.}~\bibnamefont {Vovk}}}\ (\bibinfo
   {publisher} {Springer-Verlag Berlin Heidelberg},\ \bibinfo {year} {2013})\
  pp.\ \bibinfo {pages} {245--259}\BibitemShut {NoStop}%
\bibitem [{\citenamefont {Snyder}\ \emph {et~al.}(2015)\citenamefont {Snyder},
  \citenamefont {Rupp}, \citenamefont {MŸller},\ and\ \citenamefont
  {Burke}}]{QUA:QUA24937}%
  \BibitemOpen
  \bibfield  {author} {\bibinfo {author} {\bibfnamefont {J.~C.}\ \bibnamefont
  {Snyder}}, \bibinfo {author} {\bibfnamefont {M.}~\bibnamefont {Rupp}},
  \bibinfo {author} {\bibfnamefont {K.-R.}\ \bibnamefont {MŸller}}, \ and\
  \bibinfo {author} {\bibfnamefont {K.}~\bibnamefont {Burke}},\ }\href
  {\doibase 10.1002/qua.24937} {\bibfield  {journal} {\bibinfo  {journal}
  {International Journal of Quantum Chemistry}\ }\textbf {\bibinfo {volume}
  {115}},\ \bibinfo {pages} {1102} (\bibinfo {year} {2015})}\BibitemShut
  {NoStop}%
\bibitem [{\citenamefont {Hinton}\ and\ \citenamefont
  {Salakhutdinov}(2006)}]{Hinton504}%
  \BibitemOpen
  \bibfield  {author} {\bibinfo {author} {\bibfnamefont {G.~E.}\ \bibnamefont
  {Hinton}}\ and\ \bibinfo {author} {\bibfnamefont {R.~R.}\ \bibnamefont
  {Salakhutdinov}},\ }\href {\doibase 10.1126/science.1127647} {\bibfield
  {journal} {\bibinfo  {journal} {Science}\ }\textbf {\bibinfo {volume}
  {313}},\ \bibinfo {pages} {504} (\bibinfo {year} {2006})},\ \Eprint
  {http://arxiv.org/abs/http://science.sciencemag.org/content/313/5786/504.full.pdf}
  {http://science.sciencemag.org/content/313/5786/504.full.pdf} \BibitemShut
  {NoStop}%
\end{thebibliography}%


\begin{acknowledgments}
J.\ M.\ M.\ acknowledges startup support from Washington State University and the Department of Physics and Astronomy thereat.
\end{acknowledgments}







\end{document}